\documentclass[11pt]{article}

\usepackage{color}

\definecolor{indigo}{RGB}{0,0,120}
\usepackage[colorlinks=true, linkcolor=indigo, citecolor=blue, urlcolor=indigo, backref=page]{hyperref}

\def\modulo{{\rm mod}\:}

\def\imply{\Rightarrow}

\def\fl{\noindent}

\newcommand{\tl}[1]{\tilde{#1}}

\newcommand{\dd}[2]{\frac {\partial #1}{\partial #2}}

\newcommand{\pdr}{\partial}

\newcommand{\grad}{{\bm \nabla}}

\newcommand{\beq}{\begin{equation}}
\newcommand{\eeq}{\end{equation}}
\newcommand{\beqs}{\begin{eqnarray}}
\newcommand{\eeqs}{\end{eqnarray}}

\def\al{\alpha}		\def\g{\gamma} 		\def\G{\Gamma} 
\def\del{\delta}			\def\eps{\epsilon} 
			 
		\def\sig{\sigma}		
		\def\tht{\theta}	
				\def\Om{\Omega}

\usepackage{amsmath,amssymb} % for AMS symbols and macros

\usepackage[pdftex]{graphicx}

\usepackage{calligra} % for script and cursive characters like script r
\DeclareMathAlphabet{\mathcalligra}{T1}{calligra}{m}{n}
\DeclareFontShape{T1}{calligra}{m}{n}{<->s*[2.2]callig15}{}

\newcount\colveccount  % column vec \colvec{nrows}{a11 & a12}{a21 & a22}
\newcommand*\colvec[1]{\global\colveccount#1  \begin{pmatrix} \colvecnext} \def\colvecnext#1{#1 \global\advance\colveccount-1
        \ifnum\colveccount>0 \\ \expandafter\colvecnext
        \else \end{pmatrix} \fi}

\usepackage[skip=3pt,labelfont=normal,labelsep=colon]{caption}
\usepackage[skip=3pt]{subcaption} %skip decides the spacing between subfigures and subcaptions, default=6pt.%If you want (a) to be in bold face: labelfont=bf. %default labelsep=space.
\usepackage{wrapfig}

\usepackage{cancel}
\usepackage[normalem]{ulem}
\usepackage{bm}

\newcommand{\bmv}{{\bm v}}
\newcommand{\bmw}{{\bm w}}

\begin{document}

%------------

\title{
%\hfill{\tt \small \href{https://arxiv.org/abs/24??.?????}{arXiv:24??.????? [??]}}\\
Level crossing instabilities in inviscid isothermal compressible Couette flow 
\author{{\sc Govind S. Krishnaswami, Sonakshi Sachdev and Pritish Sinha}
\\ \small
Physics Department, Chennai Mathematical Institute,  SIPCOT IT Park, Siruseri 603103, India\\
\\ \small
Email: {\tt govind@cmi.ac.in, sonakshi@cmi.ac.in, pritish@cmi.ac.in}}
\date{30 December, 2024\\ 
\vspace{0.1cm}
}}

\maketitle

% abstract is 215 words, 1441 characters

\abstract{ \small We study the linear stability of inviscid steady parallel flow of an ideal gas in a channel of finite width. Compressible isothermal two-dimensional monochromatic perturbations are considered. The eigenvalue problem governing density and velocity perturbations is a compressible version of Rayleigh's equation and involves two parameters: a flow Mach number $M$ and the perturbation wavenumber $k$. For an odd background velocity profile, there is a $\mathbb{Z}_2 \times \mathbb{Z}_2$ symmetry and growth rates $\g$ come in symmetrically placed 4-tuples in the complex eigenplane. Specializing to uniform background vorticity Couette flow, we find an infinite tower of noninflectional eigenmodes and derive stability theorems and bounds on growth rates. We show that eigenmodes are neutrally stable for small $k$ and small $M$ but that they otherwise display an infinite sequence of stability transitions with increasing $k$ or $M$. Using a search algorithm based on the Fredholm alternative, we find that the transitions are associated to level crossings between neighboring eigenmodes. Repeated level crossings result in windows of instability. For a given eigenmode, they are arranged in a zebra-like striped pattern on the $k$-$M$ plane. A canonical square-root power law form for $\g(k,M)$ in the vicinity of a stability transition is identified. In addition to the discrete spectrum, we find a continuous spectrum of eigenmodes that are always neutrally stable but fail to be smooth across critical layers. }

%{\fl \bf Keywords:} 

{\footnotesize \tableofcontents}

\normalsize

%-------------
\section{Introduction}
\label{s:intro}
%-------------

The stability of parallel shear flows has been a topic of long-standing interest, going at least as far back as the work of Kelvin and Helmholtz \cite{drazin-reid}. They are of relevance to both laboratory \cite{blumen-1975} and astrophysical flows \cite{choudhr-lov}. The roll-up of a vortex sheet is a striking physical manifestation of an instability in a shear flow. Perhaps the earliest mathematical model for shear flow instabilities was that of Rayleigh \cite{rayleigh-sound}, who considered incompressible two-dimensional monochromatic perturbations to steady inviscid parallel flow and showed that an inflection point in the background velocity profile $u(y)$ is necessary for a linear instability. Fj{\o}rtoft \cite{drazin-reid}, obtained a global necessary condition for the existence of a nonneutral mode in a   monotonic velocity profile: $u''(y) [u(y) - u(y_s)] < 0$ for all $y$, where $y_s$ is the inflection point. Howard's semicircle theorem \cite{howard-1961, draz-howard-hydro-stab-parall-invi-flo} gives a bound on the growth rate $\gamma$ for unstable eigenmodes in incompressible shear flows. Extensions of Rayleigh's stability criterion and the semicircle theorem to compressible adiabatic flows were obtained by Lees and Lin \cite{Lees-n-Lin} and Eckart \cite{Eckart} respectively. Additional results on the stability of compressible inviscid shear flows may be found in \cite{Shivamoggi-1977, Shivamoggi:comp, stab-non-hom-1, subb-jain, stab-non-hom, lin-inst-compr-flow}. In the viscous case, the story begins with the Orr-Sommerfeld equation \cite{drazin-reid} for the stability of incompressible parallel flows followed by extensive studies of the stability of boundary layer flows \cite{mack}. An account of the stability of parallel gas flows may be found in the book by Shivamoggi \cite{shivamoggi-book}. 

The present work is a detailed examination of the linear stability of an inviscid linear shear layer ($u(y) \propto y$) in the isothermal dynamics of an ideal gas confined to a channel of finite width with impenetrable walls at $y = \pm L$. From a theoretical standpoint, an inviscid Couette flow is interesting for several reasons. To begin with, a linear background velocity profile has no inflection point. In fact, it is well-known \cite{drazin-reid, rayleigh-sound, CaseKM} that inviscid incompressible Couette flow has no unstable modes: the discrete spectrum is empty while the continuous spectrum of eigenperturbations has purely imaginary growth rates. The absence of inflectional instabilities makes Couette flow a relatively uncluttered arena to study the effects of compressibility. In \cite{Glat-WV, Duck}, the stability of compressible Couette flow has been examined, although the focus is on the viscous case. In keeping with our theme of simplicity, we restrict ourselves to isothermal inviscid flow of an ideal gas, so that pressure and density are not independent and there is no need for an energy equation to supplement the continuity and compressible Euler equations. There are physical limitations that arise from our simplifications. For the atmosphere modeled as an ideal gas, the isothermal approximation fails especially at high Mach numbers, where the assumption of an adiabatic flow is more accurate. What is more, at high Mach numbers, transverse sound waves could lead to shocks, which we do not consider in our linear stability analysis. Despite these limitations, its simplicity allows us to view inviscid isothermal Couette flow as a laboratory to find and study compressional instabilities that may be harder to isolate in other flows. In fact, upon including compressible perturbations, we find that a discrete spectrum of eigenmodes germinates and displays a striking pattern of instabilities, while the continuous spectrum remains neutrally stable. It is natural to examine the growth rates $\g$ of modes as a function of the two dimensionless parameters in the problem: a perturbation wavenumber $k$ (in units of $1/L$) and a flow Mach number $M$. Using a combination of stability theorems, growth rate bounds, (asymptotic) analytical methods and a powerful numerical search algorithm, we obtain a quantitative picture of the pattern of instabilities. While the flow is neutrally stable when either $k$ or $M$ is small, each mode undergoes an infinite sequence of stability-instability transitions as $k$ or $M$ is increased. These result in bands of instability that form a zebra-like pattern (Fig.~\ref{f:instab-stripe-grnd-mode}) on the $k$-$M$ parameter plane. Interestingly, these widows of instability arise from `level crossings' (sometimes called `resonant interactions' or `mode conversions') between the growth rates of adjacent modes. As $k$ or $M$ is increased, a pair of imaginary growth rates collide on the imaginary $\g$ axis, perform symmetrical excursions into the complex plane and then merge on the same axis. Our level crossing instability transitions may be viewed as degenerate cases of transcritical bifurcations: a pair of neutrally stable modes interact leaving one stable and the other unstable. Since the tower of modes that we find arises due to compressibility, they are expelled from the spectrum as $M \to 0$, where our equations reduce to Rayleigh's eigenvalue problem \cite{rayleigh-sound}. Moreover, the instabilities we find are relatively short wavelength instabilities in contrast with Rayleigh's inflectional modes \cite{rayl-1880} and the Tollmien-Schlichting waves \cite{drazin-reid, mack} in viscous shear flows, which display long wavelength instabilities.

Although our phase diagram of instabilities as well as asymptotic approximations, a priori bounds and stability criteria seem to be new results on inviscid isothermal compressible Couette flow, some ingredients in this picture have been glimpsed by other authors. Notably, while studying sonic instabilities in an inviscid adiabatic linear supersonic shear layer in an astrophysical context, Glatzel \cite{Glat-W} found a pattern of instabilities as a function of Mach number similar to the ones we find as a function of {\it both} $M$ and $k$. However, while Glatzel's perturbations were adiabatic and homentropic, ours are isothermal. Moreover, our perturbed flows can have any specific heat ratio $\g = c_p/c_v > 1$ and are not obtainable as $\g \to 1$ limits of adiabatic flows. For recent work on the singular nature of the $\g \to 1$ limit in a polytropic gas model of the solar wind, see \cite{pohl-shivamoggi}. Furthermore, the numerical method we adopt is altogether different from the techniques used in \cite{Glat-W}. On the other hand, Renardy \cite{rena-m} provides a proof of an instability arising from eigenvalue crossing in a transonic linear shear flow in shallow water (with Froude number playing the role of Mach number) and in plane Couette flow of a viscoelastic fluid. Taken together, these results suggest that the compressional modes and level crossing instabilities that we uncover are a general feature and not artifacts of our simplifying assumptions of isothermal inviscid flow.

The picture that emerges from our linear stability analysis could be modified by effects not considered. For instance, nonlinear saturation could mute the growth of some of our linear instabilities while a linearly stable flow can be nonlinearly destabilized (e.g., via subcritical bifurcations). On the other hand, in regimes (such as those of small $k$ or $M$) where we have found neutral linear stability, there could be transient instabilities due to polynomial-time growth of generalized eigenvectors, that is not captured by eigenvectors with exponential $(e^{\g t})$ time-dependence. It is recognized \cite{trefethen-el-al} that such transients can arise when the linearized flow is described by a nonnormal operator that fails to be completely diagonalizable. See \cite{Anton-Dol-Marc} and references therein for recent work on Couette flow where such possibilities are explored.

Finally, we point out the difference between our setup and the one employed in discussions of `shear sheltering' \cite{hunt-durbin-shear-shelter-1999, thyagaraja-loureiro-knight-2002}. In the latter, a spatially localized pressure pulse is introduced in the vicinity of a point $(0,y_0)$ in a sheared flow. According to a stationary observer, the sheared flow will tend to longitudinally stretch and carry away the pulse aside from its spread due to the backward propagating wave and waves possibly reflected off upper and lower walls. Thus, it is heuristically argued that at the stationary observer's position, there cannot be any perpetual growth in the wave due to sound. By contrast, we have considered a mode of definite wave number, which is consequently not spatially localized. Moreover, we have only allowed for modes with exponential time-dependence and not addressed the full-fledged initial value problem. Thus, arguments based on shear sheltering do not contradict the linear instabilities that we have found. We now outline the organization of this paper and summarize our results.

%-------------
\section{Summary of results}
%-------------

We begin in \S \ref{s:comp-euler} by considering the inviscid isothermal dynamics of an ideal gas in a channel with impenetrable boundaries at $y = \pm L$. We take as our background, a steady constant density parallel shear flow $\bmv = (u(y),0,0)$ that is translation-invariant in the $x$-$z$ plane. Special cases include inviscid Couette flow as well as a vortex sheet in the $x$-$z$ plane. In \S \ref{s:perturbed-R-Euler-eq}, we derive linear equations for two-dimensional perturbations of the form $\hat \rho(y) e^{i k x + \g t}$ and $(\hat u(y),\hat v(y), 0) e^{i k x + \g t}$. The resulting eigenvalue problem for the growth rate $\g$ depends on two parameters: the perturbation wave number $k$ and a background flow Mach number $M$. It is a system of three coupled first order ordinary differential equations (ODEs) with coefficients depending on the undisturbed velocity profile $u(y)$. For an odd velocity profile, the absence of dissipation implies that the system admits two discrete symmetries (\S \ref{s:dissym}) that constrain eigenfunctions and ensure that eigenvalues come in 4-tuples $(\g, - \g^*, - \g, \g^*)$. The equations for 3d perturbations take a similar form and admit analogous symmetries.

In \S \ref{s:cc-flow}, we specialize to disturbances around a linear velocity profile: steady Couette flow. The eigenvalue equations are reduced in \S \ref{s:rho-u-v-w-ODE} to a single second order ODE for the vorticity (or density or velocity) perturbations with $\g$ appearing nonlinearly. Although it is transformable into a confluent hypergeometric equation (Appendix \ref{s:hyper-geo}), we do not attempt an explicit analytic determination of the spectrum of growth rates in this paper. Instead, we develop a quantitative picture of linear instabilities by combining stability theorems, growth rate bounds, limiting cases, series solutions and a numerical approach. Elsewhere, we hope to prove the key features of our numerical spectra by careful examination of the complex analytic properties of the associated ODEs and their Wronskians. In a slightly different direction, it is worth noting that in Chapt. 7 of \cite{shivamoggi-book}, Shivamoggi discusses a qualitative phase plane analysis around the trivial fixed point of an equation that reduces (at constant temperature) to that for our vorticity perturbations. However, in this analysis of subsonic and supersonic disturbances, only neutral perturbations are considered and our finite channel boundary conditions are not imposed.

In \S \ref{s:stab-thrm}, we derive stability theorems for this flow: (i) necessity of a `critical layer' inside the channel for instability, (ii) an analog of Howard's semicircle theorem bounding growth rates of unstable modes, (iii) a lower bound on $|\Im \g|$ for neutrally stable modes which disallows level crossings between conjugate modes and (iv) a necessary condition on $M$ and $k$ for a neutrally stable mode to have a critical layer within the channel. In \S \ref{s:small-k-m-CC}, the perturbation ODEs are solved analytically (with details in Appendix \ref{s:power-series-w}) in tractable limits such as small $k$, small $M$ and large oscillation frequency $|\Im \g|$, to reveal an infinite tower ($n = 0,1,2, \ldots$) of noninflectional compressional eigenmodes (and their conjugates $0^*, 1^*, \ldots$) that are neutrally stable. We find that sufficiently subsonic Couette flow ($M \ll 1$) is neutrally stable to perturbations of any wave number. At higher Mach numbers, it continues to be neutrally stable to perturbations of sufficiently large wavelength. Instabilities, if present at higher Mach numbers, can arise only from moderate or short wavelength perturbations. In \S \ref{s:lc-regime}, we use a numerical approach based on the Fredholm alternative (described in Appendix \S \ref{s:Fredholm}) to solve the perturbation equations around Couette flow. By using an iterative search method, we discover an infinite sequence of stability transitions in each mode as $k$ or $M$ is increased. The opening and closing of a window of instability coincides with the merger and subsequent demerger of the eigenvalue $\g_n$ with a neighboring eigenvalue. For a given eigenmode, these level-crossing instabilities produce a zebra-like pattern of bands in the $k$-$M$ plane, which we describe in \S \ref{s:instripes} for the ground ($n=0$) mode. In \S \ref{s:canon-form-eval-level-cross}, the behavior of an eigenvalue undergoing a level crossing (as $k$ or $M$ is varied) is captured by solutions of a canonical biquadratic equation that neatly incorporates both the discrete symmetries and a square-root power law behavior near a stability transition. In \S \ref{s:cont-spect}, we show that the linear stability equations also admit a continuous spectrum of neutrally stable eigenmodes. Although the discrete and continuous spectra can overlap, the eigenfunctions of the latter fail to be smooth across critical layers where the equations are singular due to isolated real zeros in the (imaginary part of the) Doppler-shifted growth rate.

%-------------
\section{Background flow and its perturbations}
\label{s:comp-euler-main}
%-------------

%-------------
\subsection{Steady parallel shear flow}
\label{s:comp-euler}
%-------------

We consider isothermal compressible flow of an ideal gas governed by the Euler and continuity equations and equation of state 
	\beq
	\dd{\bmv}{t} + \bmv \cdot \grad \bmv = -\frac{\grad p }{\rho}, \quad
	\dd{\rho}{t} + \grad \cdot ( \rho \bmv) = 0 \quad \text{and} \quad p = \rho k_B T / \mu. 
 \label{e:v-time-evol}
	\eeq
Here $T$ is the constant temperature throughout the flow and $\mu$ the molecular mass. Under isothermal conditions, it is convenient to introduce the specific Gibbs free energy $g(\rho) = (k_B T/\mu) \log(\rho/\rho_0)$ which is determined up to an additive constant by the condition $\grad g = (\grad p)/\rho$. Similarly, the specific entropy for isothermal flow of an ideal gas satisfying the caloric condition is defined up to a constant:
	\beq
	s = s_0 + (k_B/\mu) \log(\rho_0/\rho)
	\label{e:entropy-general}
	\eeq
and evolves according to $\dd{s}{t} + \bmv \cdot \grad s = (k_B/\mu) \grad \cdot \bmv$.

\begin{figure}[!h]
\centerline{\includegraphics[width=9cm]{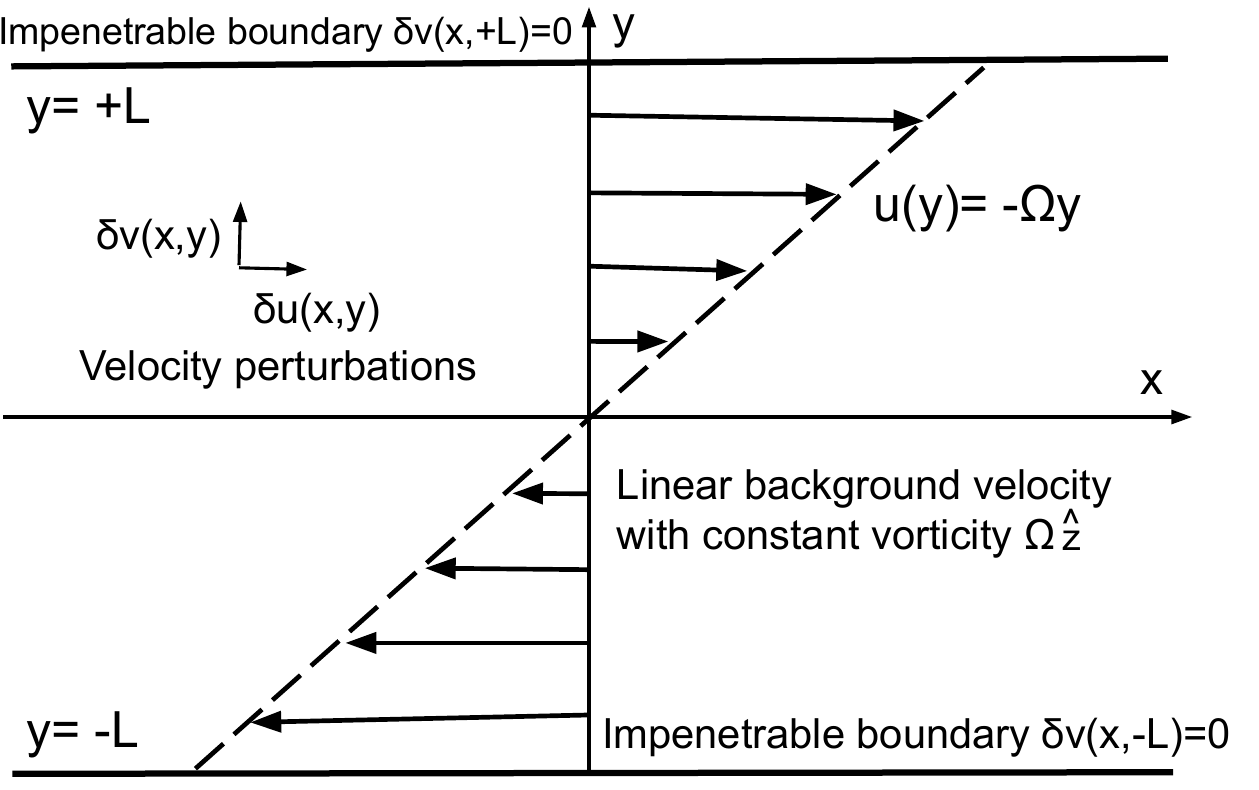}}
\caption{\small Steady inviscid plane Couette flow. The background density, pressure and temperature are uniform. Perturbations are assumed isothermal but compressible.} \label{f:vortex-sheet}
\end{figure}

We will be interested in investigating the stability of a steady plane parallel shear flow in the $x$-direction, where the density $\rho(y)$ and velocity field $(u(y), 0, 0)$ depend only on the height $y$ (see Fig.~\ref{f:vortex-sheet}). The corresponding vorticity points in the $z$ direction: $\bmw = \grad \times \bmv  = -u'(y) \hat z$. The steady continuity equation $\pdr_x(\rho(y) u(y)) = 0$ is identically satisfied. Only the $y$ component of the steady Euler equation survives, leading to $g'(y) = 0$ but leaving $u(y)$ undetermined. Thus we have a steady background flow in the streamwise $(x)$ direction with constant density and pressure:
	\beq
	\bar \bmv = (u(y), 0 ,0 ), \quad \rho = \rho_0 \quad \text{and} \quad p = p_0 = \rho_0 k_B T/\mu. 
	\label{e:stead-bag-prof}
	\eeq
As a consequence of (\ref{e:entropy-general}), this background flow is homentropic: spatially homogeneous with specific entropy $s_0$. We will be particularly interested in such a flow in a channel of finite height $|y| \leq L$. The impenetrability conditions at the channel boundaries ($y = \pm L$) are automatically satisfied by (\ref{e:stead-bag-prof}). In the absence of viscosity, the tangential velocity components $u(\pm L)$ are unconstrained.

The ratio of background flow speed $u_0$ at a reference height [e.g., $u_0 = |u(L)|$] to the isothermal sound speed $c_s$ is a useful dimensionless Mach number associated to this background flow:
	\beq
	M^2 =  u_0^2/c_s^2 = u_0^2 \mu/k_B T = {\rho_0 u_0^2}/{p_0}.
	\label{e:Mach-no}
	\eeq
From \S \ref{s:cc-flow} onwards, we will specialize to a linear background velocity profile 
	\beq
	u(y) = - \Om y \quad \text{with} \quad \rho(y) = \rho_0.
	\label{e:couette-ideal-flow}
	\eeq
The corresponding background vorticity is uniform: $\bmw = \Om \hat z$. We will use the term {\it inviscid planar Couette flow} for this undisturbed profile.

%-----------
\subsection{Compressible perturbations to undisturbed flow}
\label{s:perturbed-R-Euler-eq}
%------------

To study the linear stability of the plane shear flow (\ref{e:stead-bag-prof}), we will consider compressible isothermal perturbations
	\beq
	\bmv = \bar \bmv(y) + \del \bmv(x, y)  \quad \text{and} \quad
	\rho = \rho_0 + \del \rho(x, y).
	\eeq
The perturbations are taken to be two-dimensional since one expects, based on an extrapolation of Squire's Theorem \cite{drazin-reid}, that the least stable perturbations are two-dimensional. Thus $\del \bmv = (\del u, \del v, 0)$ and $\del \rho$ are assumed to preserve $z$-translation invariance but could depend on $x, y$ and $t$. Due to the $x$ translation-invariant background, we may take the perturbations to be monochromatic in their dependence on $x$ and thus write
	\beq
	\del \rho = \hat \rho(y) e^{\g t + i k x}, \quad
	\del u = \hat u(y) e^{\g t + i k x} \quad \text{and} \quad 
	\del v = \hat v(y) e^{\g t + i k x}.
	\label{e:perturb}
	\eeq
The growth rate $\gamma(k)$ would be purely imaginary for oscillatory modes while it would have a positive real part for unstable modes. Although $\tl k$ can have either sign we will take $\tl k \geq 0$ to avoid having to write $|\tl k|$ in various places below. The physical perturbations are given by the real parts $\Re (\hat \rho e^{\g t + i k x})$ etc., with both $\Re \hat \rho$ and $\Im \hat \rho$ contributing. Since the equations for perturbations are linear, we are free to work with complex disturbances and finally take real parts.

It is convenient to work with dimensionless variables. We use the constant background density $\rho_0$, reference flow speed $u_0$ (e.g., flow speed $|u(L)|$ at the boundary) and some reference length $\tht$ (such as the semi-height $L$ of the channel) to nondimensionalize variables:
	\beqs
	\tl y &=& \frac{y}{\tht}, \quad
	U = \frac{u}{u_0}, \quad 	 
	\tl \rho = \frac{\hat \rho}{\rho_0}, \quad 
	\tl u = \frac{\hat u}{u_0}, \quad 
	\tl v = \frac{\hat v}{u_0}, \quad 
	\tl k = \tht k, \quad
	\tl \g = \frac{\g \tht}{u_0}.
	\label{e:dim-less-var}
	\eeqs
The linearized continuity equation
	\beq
	\pdr_t \del \rho + \pdr_x (\rho \, \del u + u \, \del \rho) + \pdr_y (\rho \del v) = 0
	\eeq
becomes
	\beq
	\G \tl \rho = - (i \tl k  \tl u +  \tl v')
%	\tl \g \tl \rho + i \tl k (\tl u + U \tl \rho) +  \tl v' = 0,
	\label{e:density-pert-vortex-sheet}
	\eeq
where primes denote derivatives with respect to $\tl y$ and
	\beq
	\G(y) = \tl \g + i \tl k U.
	\label{e:Doppler-shifted-gamma}
	\eeq
is a height-dependent `Doppler-shifted' growth rate. 

The specific entropy perturbation $\del s = s - s_0 = \hat s e^{\g t + i k x}$ is nondimensionalized via $\tl s = \mu \hat s/k_B$. Linearizing (\ref{e:entropy-general}) we find that it satisfies $\G \tl s = i \tl k \tl u + \tl v'$. Thus $\tl s = - \tl \rho$. Notably, although our isothermal background flow is homentropic, the perturbations are not. This is in contrast with the homentropic adiabatic perturbations studied by Glatzel \cite{Glat-W}. We note that our perturbed isothermal flow cannot in general be obtained by taking the limit of an adiabatic flow as the specific heat ratio $c_p/c_v \to 1$. In particular, the isothermal gas flows considered in this work can have any specific heat ratio $c_p/c_v \geq 1$.

Next, we compute the $x$ and $y$ components of the terms in the linearized Euler equation
	\beq
	\del (\pdr_t \bmv) + \del(\bmv \cdot \grad \bmv) = - \del \grad g.
	\eeq
The linearized time derivative terms are
	\beq
	\del \bmv_t = \frac{u_0^2}{\tht} \tl \g e^{i k x + \g t}(\tl u, \tl v).
	\eeq
The components of the linearized advection term are
	\beq
	\del (\bmv \cdot \grad \bmv) = \frac{u_0^2}{\tht}e^{i k x + \g t} (\tl v U' + i \tl k U \tl u, i \tl k U \tl v).
	\eeq
The perturbation of the pressure gradient term is 
	\beq
	\del \grad g = \frac{p_0}{\rho_0} \grad \left( \frac{\del \rho}{\rho_0} \right) = \frac{ u_0^2}{\tht M^2} e^{i k x + \g t} (i \tl k \tl \rho, \tl \rho').
	\eeq
Thus, the $x$ and $y$ components of the linearized Euler equation are
	\beq
	\G \tl u + \tl v U' = - \frac{1}{M^2 } i \tl k \tl \rho  \quad \text{and} \quad
	\G \tl v = - \frac{1}{M^2} \tl \rho',
	\label{e:x-y-cpt-perturb-mom-eq-vortex-sheet}
	\eeq
Combining (\ref{e:density-pert-vortex-sheet}) and (\ref{e:x-y-cpt-perturb-mom-eq-vortex-sheet}), the perturbation equations are the eigenvalue problem
	\beq
	Q\colvec{3}{\tl \rho}{\tl u}{\tl v} = \tl \g \colvec{3}{\tl \rho}{\tl u}{\tl v}
	\quad \text{where} \quad
	Q = \colvec{3}{- i \tl k U & - i \tl k  & -  \pdr_{\tl y}}{ - i \tl k M^{-2}  & - i \tl k U  & -U' }{  - M^{-2}\pdr_{\tl y}   & 0  &  - i \tl k  U  }.
	\label{e:compr-cons-Orr-Somm}
	\eeq
We note in passing that even for $k = 0$, this operator $Q$ does not commute with its adjoint, relative to the $L^2$ inner product. Although we do not explore it here, this opens up the possibility of generalized eigenmodes that display transient polynomial time-dependence (see \cite{trefethen-el-al, Anton-Dol-Marc}). Returning to (\ref{e:compr-cons-Orr-Somm}), the scalar diagonal part of $Q$ may be combined with the eigenvalue $\g$ to give the Doppler-shifted growth rate $\G(y)$, in terms of which (\ref{e:compr-cons-Orr-Somm}) becomes
	\beq
	(Q + i \tl k U(\tl y)) \psi = \G(\tl y) \psi \quad
	\text{where} \quad \psi = \colvec{1}{\tl \rho & \tl u & \tl v}^t.
	\eeq
The above system of equations are a compressible analogue of Rayleigh's equation (\ref{e:rayleigh-pert}) for incompressible perturbations to constant density shear flows.

\iffalse
% see Eq. 25.12 on p.156 of Ref. \cite{drazin-reid} for incompressible shear flows. 
\fi

\paragraph{Incompressible limit and Rayleigh's equation.} In fact, the linearized equations (\ref{e:compr-cons-Orr-Somm}) simplify in the incompressible limit. To take this limit, we expand variables in powers of $M^2$ and take $M^2 \to 0$ \cite{Lagerstrom}. For the pressure gradient terms in (\ref{e:x-y-cpt-perturb-mom-eq-vortex-sheet}) to have a finite limit we must have $\tl \rho = \rho_* M^2 + O(M^4)$ while $ \tl u, \tl v, \tl \g$ and $U$ assume nonzero finite limits. The perturbation equations as $M \to 0$ are
	\beq
	\tl v' = - i k \tl u, \qquad
	\G \tl u = -  i k \rho_* - U' \tl v \qquad \text{and} \qquad
	\G \tl v = -  \rho_*'
	\eeq
where $\G = \tl \g + i \tl k U$. Eliminating $\rho_*$ and $\tl u$ from these equations we arrive at a single equation for $\tl v$ which agrees with Rayleigh's equation \cite{rayleigh-sound}:
	\beq
	\left( \frac{\g}{ik} + U \right) (v'' - k^2 v) - U'' \tl v  = 0.
	\label{e:rayleigh-pert}
	\eeq

\paragraph{Mach number for perturbation phase velocity.} Introducing $\g = \g_r + i \g_i$, the perturbations are $\sim e^{i(k x + \g_i t) + \g_r t}$. Thus, the phase velocity of perturbations is $c_p = - \g_i/k$. The nondimensional phase velocity is $\tl c_p = -\tl \g_i/\tl k = c_p/u_0$. It is convenient to introduce a Mach number for the phase speed of perturbations
	\beq
	M_p = |c_p/c_s| \quad \text{or} \quad M_p = |\tl c_p| M = |\tl \g_i M/\tl k|,
	\label{e:mach-pert}
	\eeq	
which we will refer to in the sequel. A related height-dependent Mach number is also useful, where the phase speed of the disturbance with respect to an observer moving with the background flow at height $y$ is considered: 
	\beq
	M_{\rm rel}(y) = |(\Im \G(y)) M/\tl k|.
	\label{e:rel-mach-num-y-dependent}
	\eeq

%--------------

\paragraph{Three-dimensional perturbations.} If we allow for 3d velocity perturbations whose dependence on $x$ and $z$ may be taken monochromatic, i.e.,
	\beq
	\del \rho = \hat \rho(y) e^{\g t + i (k_x x + k_z z)} \quad \text{and} \quad	
	\del \bmv = \colvec{3}{\hat v_x(y)}{\hat v_y(y)}{\hat v_z(y)}  e^{\g t + i (k_x x + k_z z)},
	\label{e:perturb-z}
	\eeq
then the perturbation equations become
	\beq
	Q\colvec{4}{\tl \rho}{\tl v_x}{\tl v_y}{\tl v_z} = \tl \g \colvec{4}{\tl \rho}{\tl v_x}{\tl v_y}{\tl v_z}
	\quad \text{where} \quad 
	Q = \colvec{4}{- i \tl k_x U & - i \tl k_x  & -  \pdr_{\tl y} & - i \tl k_z}{ - i \tl k_x M^{-2} & - i \tl k_x U  & -U' & 0 }{  - M^{-2} \pdr_{\tl y}   & 0  &  - i \tl k_x  U &  0 }{   - i \tl k_z M^{-2}  & 0  & 0 &   - i \tl k_x  U }.
	\label{e:compr-euler-3d-iso}
	\eeq
Upon putting $\tl k_z = 0$, this reduces to the 2d case (\ref{e:compr-cons-Orr-Somm}), as $\hat v_z$ must necessarily vanish.

%-------------------
\subsection{Discrete symmetries of perturbation equations}
\label{s:dissym}
%-------------------

In the absence of dissipation, ours is a conservative system. Thus, we should expect each growing mode to be paired with a decaying mode. This leads to a discrete $\mathbb{Z}_2$ symmetry of (\ref{e:compr-cons-Orr-Somm}).  If in addition, the background profile $U(y)$ is an odd function, then (\ref{e:compr-cons-Orr-Somm}) admits another $\mathbb{Z}_2$ symmetry. In particular, for Couette flow, i.e., linear $U(y)$ there is a symmetry about the center of the channel. These symmetries allow us to restrict the search for eigenvalues to one quadrant of the complex $\tl \g$ plane. They also constrain the form of eigenfunctions.

{\noindent \bf (A)} Taking the complex conjugate of (\ref{e:compr-cons-Orr-Somm}) and adjusting signs, it may be written as $Q (\tl \rho^*, \tl u^*, - \tl v^*)^t = - \tl \g^* (\tl \rho^*, \tl u^*, -\tl v^*)^t$. Thus, if $(\tl \rho, \tl u, \tl v)^t$ is an eigenvector with eigenvalue $\tl \g$, so is $(\tl \rho^*, \tl u^*, -\tl v^*)^t$ with eigenvalue $-\tl \g^*$. This symmetry relates an exponentially growing to a decaying mode, both either left- or right-moving.

{\noindent \bf (B)} There is an additional symmetry if $U(\tl y)$ is odd. Upon taking $\tl y \to -\tl y$ in (\ref{e:compr-cons-Orr-Somm}), we get $Q (\tl \rho (-\tl y), - \tl u (-\tl y), \tl v (-\tl y))^t = - \tl \g (\tl \rho (-\tl y), - \tl u (-\tl y), \tl v (-\tl y))^t$. Thus, if $(\tl \rho (\tl y), \tl u (\tl y), \tl v (\tl y))^t$ is a solution with eigenvalue $\tl \g$, then so is  $(\tl \rho (-\tl y), - \tl u (-\tl y), \tl v (-\tl y))^t$ with eigenvalue $-\tl \g$. {\bf Remarks:} (i) If $\Re \tl \g \ne 0$, this means an exponentially growing right-moving perturbation is paired with an exponentially decaying left-moving perturbation. (ii) If $\Re \tl \g = 0$, then this symmetry relates an oscillatory left-moving perturbation to an oscillatory right-moving one.

%--------------

\paragraph*{Eigenvalues come in 4-tuples.} Combining symmetries (A) and (B) which comprise the group $\mathbb{Z}_2 \times \mathbb{Z}_2$, we see that eigenmodes generically come in 4-tuples. If any one of the following is a solution to (\ref{e:compr-cons-Orr-Somm}), then so are the other three:
	\beqs
	(\tl \rho (\tl y), \tl u (\tl y), \tl v (\tl y), \tl \g), &&
	(\tl \rho^* (\tl y), \tl u^* (\tl y), -\tl v^* (\tl y), -\tl \g^*), \cr
	(\tl \rho (-\tl y), -\tl u (-\tl y), \tl v (-\tl y), -\tl \g) & \text{and} &
	(\tl \rho^* (-\tl y), -\tl u^* (-\tl y), -\tl v^* (-\tl y), \tl \g^*).
	\label{e:4-tuples}
	\eeqs

%-----------------------------

\paragraph*{Forms of eigenfunctions for real or imaginary $\tl \g$.} Assuming nondegeneracy of eigenvalues, these symmetries restrict eigenfunctions. E.g., for real $\tl \g$ and odd background velocity $U(\tl y)$, $(\tl \rho, \tl u, \tl v)$ must be of the form $e^{i \al} (\rho_e(\tl y)+ i \rho_o(\tl y), u_o(\tl y)+ i u_e(\tl y), v_o(\tl y)+ i v_e(\tl y))$, where the subscripts $e$ and $o$ denote even and odd functions and $\al$ is real. Similarly, for imaginary $\tl \g$, $(\tl \rho, \tl u, \tl v)$ must be of the form $e^{i \al}(\rho_r, u_r, i v_i)$ where $\rho_r, u_r$ and $v_i$ are real functions of $\tl y$.  We now show how these arise.

\vspace{5pt}

{\fl \bf Real $\tl \g$.} Applying both symmetries, if $(\tl \g, \tl \rho (\tl y), \tl u (\tl y), \tl v (\tl y))$ is a solution, so is $(\tl \g^*, \tl \rho^* (-\tl y), -\tl u^* (-\tl y), - \tl v^* (- \tl y))$. Now, if $\tl \g$ is real and nondegenerate, then $(\tl \rho (\tl y)$, $\tl u (\tl y)$, $\tl v (\tl y))$  must be a multiple of $(\tl \rho^* (-\tl y), -\tl u^* (-\tl y), - \tl v^* (- \tl y))$. The multiplicative factor must be a phase $e^{i \tht}$ as the symmetries do not alter the magnitudes of any of the fields. Isolating the real and imaginary parts, we get the following condition:
	\beqs
	\colvec{3}{\tl \rho_r (\tl y)+ i \tl \rho_i (\tl y)}{\tl u_r (\tl y)+ i \tl u_i (\tl y)}{\tl v_r (\tl y)+ i \tl v_i (\tl y)} 
	= e^{i \tht}\colvec{3}{\tl \rho_r (-\tl y)- i \tl \rho_i (-\tl y)}{-\tl u_r (-\tl y)+ i \tl u_i (-\tl y) }{-\tl v_r (-\tl y)+ i \tl v_i (-\tl y)} \quad \text{for all $\tl y$}.
	\label{e:symm-1} 
	\eeqs
In particular, if $\theta = 0$ then $\tl \rho_r$, $\tl u_i$ and $\tl v_i$ must be even and $\tl \rho_i$, $\tl u_r$ and $\tl v_r$ must be odd. More generally, we will show that $\tl \rho = e^{i \al} (\rho_e + i \rho_o)$ while $\tl u = e^{i \al} (u_o + i u_e)$ and $\tl v = e^{i \al} (v_o + i v_e)$ where $\al \equiv \tht/2 \;\; \modulo \pi$ and the subscripts $e, o$ denote even and odd real functions of $\tl y$. To see this, we rewrite (\ref{e:symm-1}) as
	\beq
	e^{-i \tht/2}\colvec{3}{\tl \rho (\tl y)}{\tl u (\tl y)}{\tl v (\tl y)} = e^{i \tht/2} \colvec{3}{\tl \rho^* (-\tl y)}{-\tl u^* (-\tl y)}{-\tl v^* (-\tl y)} 
	\quad \Leftrightarrow \quad
	e^{-i \tht/2}\colvec{3}{\tl \rho (\tl y)}{\tl u (\tl y)}{\tl v (\tl y)} 
	= \colvec{3}{e^{-i \tht/2} \tl \rho(-\tl y)}{-e^{-i \tht/2} \tl u(-\tl y)}{-e^{-i \tht/2} \tl v (-\tl y)}^* 
	\label{e:odd-even-yet-real-eg}
	\eeq
Thus $\chi = e^{-i \tht/2}(\tl \rho (\tl y), \tl u (\tl y), \tl v (\tl y))$ satisfies (\ref{e:symm-1}) with $\tht = 0$. Consequently, $\Re \chi_1$, $\Im \chi_2$ and $\Im \chi_3$ must be even while $\Im \chi_1$, $\Re \chi_2$ and $\Re \chi_3$ must be odd, leading to:
	\beq
	\tl \rho = e^{i \al} (\rho_e(\tl y)+ i \rho_o(\tl y)), \quad
	\tl u = e^{i \al} (u_o(\tl y)+ i u_e(\tl y)), \quad
	\tl v = e^{i \al} (v_o(\tl y)+ i v_e(\tl y)).
	\eeq
Here $\alpha \equiv \theta/2$ $(\modulo \pi)$ where we have included a possible overall minus sign in defining the functions on the RHS.

\vspace{5pt}

{\fl \bf Imaginary $\tl \g$.} A similar analysis applies when $\tl \g$ is imaginary. Under symmetry (A), the eigenpair $(\tl \g, \tl \rho (\tl y), \tl u (\tl y), \tl v (\tl y))$ $\mapsto $ $(-\tl \g^*, \tl \rho^* (\tl y), \tl u^* (\tl y), - \tl v^* (\tl y))$. If $\tl \g$ is imaginary and nondegenerate, then as before, $(\tl \rho (\tl y), \tl u (\tl y),  \tl v (\tl y))$ can differ from $(\tl \rho^* (\tl y)$, $\tl u^* (\tl y)$, $- \tl v^*(\tl y))$ at most by a multiplicative phase:
	\beq
	\colvec{3}{\tl \rho_r (\tl y)+ i \tl \rho_i (\tl y)}{\tl u_r (\tl y)+ i \tl u_i (\tl y)}{\tl v_r (\tl y)+ i \tl v_i (\tl y)} 
	= e^{i \tht}\colvec{3}{\tl \rho_r (\tl y)- i \tl \rho_i (\tl y)}{\tl u_r (\tl y)- i \tl u_i (\tl y) }{-\tl v_r (\tl y)+ i \tl v_i (\tl y)} \quad \text{for all $\tl y$}.
	\label{e:real-imag-imag-eg}
	\eeq
If $\theta = 0$, $\tl \rho, \tl u$ must be real and $\tl v$ must be imaginary. More generally, $\tl \rho$ and $\tl u$ can differ from real functions and $\tl v$ can differ from an imaginary function at most by a phase: $\tl \rho(\tl y) = e^{i \al} \tl \rho_r (\tl y)$, $\tl u(\tl y) = e^{i \al} \tl u_r (\tl y)$ and $\tl v(\tl y) = e^{i \al} i \tl v_i (\tl y)$ where $\alpha=\tht/2$ $(\modulo \pi)$. To see this, we rewrite (\ref{e:real-imag-imag-eg}) as
	\beq
	e^{-i \tht/2}\colvec{3}{\tl \rho (\tl y)}{\tl u (\tl y)}{\tl v (\tl y)} = e^{i \tht/2} \colvec{3}{\tl \rho^* (\tl y)}{\tl u^* (\tl y)}{-\tl v^* (-\tl y)} 
	\quad \Leftrightarrow \quad
	e^{-i \tht/2}\colvec{3}{\tl \rho (\tl y)}{\tl u (\tl y)}{\tl v (\tl y)} 
	= \colvec{3}{e^{-i \tht/2} \tl \rho(\tl y)}{e^{-i \tht/2} \tl u(\tl y)}{-e^{-i \tht/2} \tl v (\tl y)}^*.
	\label{e:odd-even-yet-imag-eg}
	\eeq
Thus, $e^{-i \tht/2} \tl \rho(\tl y)$ and $e^{-i \tht/2} \tl u (\tl y)$ must be real and $e^{-i \tht/2} \tl v (\tl y)$ must be imaginary, leading us to the proposed forms.

\paragraph{Symmetries of 3d perturbation equations.} Symmetries (A) and (B) also apply to 3d perturbations (\ref{e:compr-euler-3d-iso}). In particular, eigenmodes continue to come in 4-tuples: \small
	\beqs
	(\tl \rho (\tl y), \tl v_x (\tl y), \tl v_y (\tl y), \tl v_z (\tl y), \tl \g), &&
	(\tl \rho^* (\tl y), \tl v^*_x (\tl y), -\tl v^*_y (\tl y), \tl v^*_z (\tl y), -\tl \g^*), \cr
	(\tl \rho (-\tl y), -\tl v_x (-\tl y), \tl v_y (-\tl y), -\tl v_z (-\tl y), -\tl \g) & \text{\&} &
	(\tl \rho^* (-\tl y), -\tl v^*_x (-\tl y), -\tl v^*_y (-\tl y), -\tl v^*_z (-\tl y), \tl \g^*). \quad
	\label{e:4-tuples-3d} \;\;
	\eeqs \normalsize
As a consequence of these symmetries, the eigenfunctions must take certain simple canonical forms if the eigenvalues are either real or purely imaginary. For a nondegenerate real eigenvalue $\tl \g = \tl \g^*$, the eigenfunction $(\tl \rho, \tl v_x , \tl v_y , \tl v_z)$ is always of the form $e^{i \al} (\tl \rho_e + i \tl \rho_o, \tl v_{xo} + i \tl v_{xe}, \tl v_{yo} + i \tl v_{ye}, \tl v_{zo} + i \tl v_{ze})$. Similarly, for a nondegenerate imaginary eigenvalue $\tl \g = - \tl \g^*$, the eigenfunction must be of the form $e^{i \al}(\tl \rho_r, \tl v_{xr}, i \tl v_{yi}, \tl v_{zr})$.

%------------
\section{Stability of compressible Couette flow}
\label{s:cc-flow}
%------------

%------------
\subsection{Perturbation equations for linear background velocity}
\label{s:rho-u-v-w-ODE}
%------------

In this section, we focus on velocity perturbations ($\tl u (\tl y), \tl v (\tl y)$) around a linear background velocity ($u( y) = -\Om y$) corresponding to a constant background vorticity ($w_z(y) = - u'(y) = \Om$). Thus, we will be studying the linear stability of isothermal compressible planar inviscid Couette flow of an ideal gas. We focus on  flow in a layer with impenetrable boundaries at $y = \pm L$ (so that the perturbation $\hat{v} (y = \pm L) = 0$). Now, we choose our reference values (\ref{e:dim-less-var}) for nondimensionalization as:
	\beqs
	\tht = L \quad \text{and} \quad
	u_0 = - \Om L, \quad
	\text{so that} \quad U'(\tl y) = 1.
	\label{e:ref-val}
    \eeqs
The Mach number $M$ defined in (\ref{e:Mach-no}) then becomes
	\beqs
	M^2 =  \frac { u_0^2}{c_s^2} = \frac{\Om^2 L^2}{k_B T/\mu}.
	\label{e:ref-val-M} 
    \eeqs
Note that this is the {\it maximum} Mach number of the Couette flow. The nondimensional perturbation equations (\ref{e:density-pert-vortex-sheet}) and (\ref{e:x-y-cpt-perturb-mom-eq-vortex-sheet}) for the linear background $U(\tl y) = \tl y$ are
	\beq
    \label{cons-pert}
	\G \tl \rho = -  (i \tl k \tl{u}+\tl{v}'), \quad
 	\G \tl u + \tl v  =- i \tl k \frac{ \tl\rho}{M^2 }   
	\quad \text{and} \quad
     \G \tl v = - \frac{ \tl\rho'}{M^2 }
    \eeq
where $\G  = \tl \g + i \tl k \tl y$ is linear in $\tl y$. We will now show that these coupled equations may be reduced to self-contained $2^{\rm nd}$ order ODEs for any one of the disturbances: $\tl u$, $\tl v$, $\tl \rho$ or the vorticity perturbation $\tl w = i \tl k \tl v - \tl u'$ with appropriate boundary conditions (BCs): Dirichlet for $\tl v$, Neumann for $\tl w$ and $\tl \rho$  and Robin for $\tl u$. Subject to these BCs, we may of course rescale the eigenmode $(\tl \rho, \tl u, \tl v)$ corresponding to eigenvalue $\tl \g$ by an overall normalization factor. Interestingly, the BC on $\tl v$ makes it more convenient for numerical purposes (see \S \ref{s:lc-regime})  while the equation for $\tl w$ is suited for the analytic approaches in \S \ref{s:small-k-m-CC} and Appendix \ref{s:power-series-w}.

\paragraph{Equation for $\tl w$.} Taking the `curl' $i \tl k(y-{\rm cpt})-(x-{\rm cpt})'$ and using the $\tl \rho$ equation, we find that $\tl w = - \tl \rho$:
	\beqs
	i \tl{k} \G \tl v - (\G \tl{u})' + (\tl v \G)' = \G(i \tl k \tl v - \tl u') -(i \tl k \tl u + \tl v') = 0 \quad
	\imply \quad \tl w =  - \tl \rho.
  \label{e:vort-rho-pert}
  \eeqs
Thus, we eliminate $\tl \rho$ in favor of $\tl w$ in the perturbation equations (\ref{cons-pert}):
	\beq 
	\G \tl w =   (i \tl k \tl{u}+\tl{v}'), \quad
\Gamma \Tilde{u} + \Tilde{v} =  \frac{i \tl k}{M^2}  \tl w   \quad
	\text{and} \quad \G \Tilde{v} = \frac{1}{M^2}\tl w' .
	\label{u-v-w}
    \eeq
The latter two equations allow us to write $\tl u$ and $\tl v$ in terms of $\tl w$:
	\beq
	\tl u =  \frac{1}{\G M^2} \left(i \tl k \tl w -\frac{\tl w'}{\G}  \right)
	\quad \text{and} \quad 
	\tl v = \frac{\tl w'}{\G M^2}.
	\label{u-v-w-P}
	\eeq
Putting (\ref{u-v-w-P}) in the first equation of (\ref{u-v-w}), we get a self-contained equation for $\tl w$:
	\beq
	\tl w''-2\frac{i  \tl k }{\G} \tl w' - ({\tl k}^2+M^2\G^2) \tl w = 0.
	\label{w-diff-eqn}
	\eeq
To determine $\tl w$ from the `eigenvalue' problem (\ref{w-diff-eqn}), we must supplement the BCs $\tl w'(\pm 1) = 0$ (coming from the second equation of (\ref{u-v-w-P})) with, say, the value of $\tl w$ somewhere to set the scale as this is a linear system. This equation has a regular singularity at $\tl y_c =  -\tl \g/i \tl k $ where $\G$ vanishes and an essential singularity at $\tl y = \infty$. The singularity at $\tl y_c $ is related to the concept of a critical layer, which will be discussed in \S \ref{s:stab-thrm}. Having found $\tl w$, we may use (\ref{e:vort-rho-pert}) and (\ref{u-v-w-P}) to get $\tl \rho, \tl u$ and $\tl v$. In fact, $\tl \rho$ satisfies the same equation as $\tl w$.

As shown in Appendix \ref{s:hyper-geo}, Eqn. (\ref{w-diff-eqn}) can be transformed into the confluent hypergeometric equation as Glatzel \cite{Glat-W} does in his study of the stability of Couette flow under adiabatic conditions (see also the earlier work of Dyson \cite{dyson}). The allowed eigenvalues $\tl \g$ are determined by imposing boundary conditions leading to the transcendental equation (\ref{e:bc-hyp}). Since an explicit solution of (\ref{e:bc-hyp}) is not available, we will resort to asymptotic approximations and numerical approaches to find the spectrum of growth rates in \S \ref{s:small-k-m-CC} and \S \ref{s:lc-regime}.

\paragraph{Remark.} It is noteworthy that (\ref{w-diff-eqn}) also arises in Shivamoggi's discussion of subsonic and supersonic inviscid pressure disturbances in parallel flows. Indeed Eqn. 10 of Chapt. 7 of \cite{shivamoggi-book} for pressure perturbations at a constant temperature reduces to our vorticity perturbation equation (\ref{w-diff-eqn}). In this reference, a qualitative phase plane analysis of this nonautonomous equation is discussed around the fixed point $(\tl w = 0, \tl v = 0)$. It reveals a distinction between subsonic and supersonic disturbances: a change from saddle to node. The disturbances are classified as subsonic and supersonic using $M_{\rm rel}$ of (\ref{e:rel-mach-num-y-dependent}). However, some differences from our study must be pointed out. In the analysis of \cite{shivamoggi-book}, $\tl \g$ is imaginary so only neutral disturbances are considered. More importantly, the channel is unbounded, so our boundary conditions at $y = \pm L$ are not imposed. Thus, our discrete spectrum of modes does not appear in this phase plane analysis.

%--------------------------- 
\paragraph{Equation for $\tl v$.} Given that $\tl v$ satisfies Dirichlet BCs, it is advantageous (especially for numerical purposes) to have an ODE for $\tl v$. Differentiating the equation for $\tl v$ in (\ref{u-v-w-P}), we get 
	\beq
	\G \tl v' + i \tl k \tl v = \frac{1}{M^2} \tl w''.
	\label{v-vpr-wprpr-rel}
	\eeq
We use (\ref{u-v-w-P}) and (\ref{v-vpr-wprpr-rel}) in (\ref{w-diff-eqn}) to eliminate $\tl w'$ and $\tl w''$ in favor of $\tl v$ and $\tl v'$
	\beq
	\G\tl v' - i \tl k \tl v = \bigg(\frac{{\tl k}^2}{M^2}+\G^2\bigg) \tl w.
	\label{v-diff-eqn}
	\eeq
Differentiating in $\tl y$ and using $\G' =  i \tl k$, we get 
	\beq
	\G \tl v'' = \left( \frac{\tl k^2}{M^2}+\G^2 \right) \tl w' + 2 \G i \tl k  \tl w .
	\label{e:v''-w-w'}
	\eeq
Finally, we use (\ref{u-v-w-P}) and (\ref{v-diff-eqn}) to arrive at an ODE for $\tl v$:
	\beq
	(\tl k^2 + M^2 \G^2) \tl v'' - 2 M^2\G i \tl k  \tl v' - ((\tl k^2 + M^2 \G^2)^2 + 2 \tl k^2 M^2) \tl v = 0.
	\label{e:v-diff-eq-const-vortex}
	\eeq
This equation has 2 regular singular points at $\tl y = \pm 1/M + i \tl \g/\tl k$ and one essential singularity at $\tl y = \infty$. Having found $\tl v$, we may get $\tl w = - \tl \rho$ by using (\ref{v-diff-eqn}). Finally, $\tl u$ can be obtained from the algebraic relation $\G \tl{\rho} = -  (i \tl k\tl u + \tl v')$ (\ref{cons-pert}) or from (\ref{u-v-w-P}).

\paragraph{Equation for $\tl u$.} The equation for $\tl u$ is structurally the simplest although it satisfies Robin BCs. We use $\tl v = \tl w'/(\G M^2)$ (\ref{u-v-w-P}) to eliminate $\tl v$ from $\tl w = i \tl k \tl v - \tl u'$ to get $\G \tl u' = \G \tl w - i \tl k \tl w'/M^2$. We also have $M^2\G^2 \tl u =  \G i \tl k \tl w - \tl w'$ from (\ref{u-v-w-P}). Combining, we write $\tl w$ and $\tl w'$ in terms of $\tl u$ and $\tl u'$:
	\beq
	\tl w = -M^2\frac{( \tl u' + i \tl k \G \tl u )}{(\tl k^2  + M^2)} \quad 
	\text{and} \quad \tl w' = -\G M^2 \frac{(i \tl k \tl u' + M^2\G \tl u )}{(\tl k^2 + M^2)}.
	\label{w-and-wpr-intermsof-u} 
	\eeq
Equating expressions for $w'$, we get an ODE for $\tl u$:
	\beq
\tl u'' - (\tl k^2 + M^2 \G^2 )\tl u = 0.
	\label{u-2nd-order-ode} 
	\eeq
It has no regular singular points but one essential singularity at $\tl y = \infty$. The BCs $\tl w' (\pm 1) = 0$ upon use of (\ref{w-and-wpr-intermsof-u}) take a Robin form: $i \tl k \tl u' (\pm 1) + (\tl \g \pm i \tl k) M^2 \tl u (\pm 1)  = 0$, provided $\G(\pm 1) \neq 0$.

%-----------------------
\subsection{Stability theorems: bounds and instability criteria} 
\label{s:stab-thrm}
%-----------------------

Here, we derive conditions for the background Couette flow to be stable/unstable.  (i)  We show that $|\Im \tl \g| < \tl k$ is a necessary condition for a mode to be unstable. Writing $\tl \g = \tl \g_r + i \tl \g_i$, this is the condition that the phase velocity $c = -\tl \g_i/\tl k$ of the mode matches that of the background flow $U(\tl y) = \tl y$ at some $\tl y = \tl y_c$ in the channel $|\tl y| \leq 1$. Such a $\tl y_c$, which may also be defined by the condition $\Im \G(\tl y_c) = \g_i + \tl k \tl y_c = 0$, is called a critical layer\footnote{Some authors (see \S \ref{s:cont-spect}) define a critical layer by $\G(\tl y_c) = 0$ rather than $\Im \G(\tl y_c) = 0$. By this definition, since $\G = \tl \g + i \tl k \tl y$, the `critical layer' $\tl y_c$ of an unstable mode would not be real although $\Re \tl y_c$ would correspond to our critical layer lying inside the channel.}. So a critical layer must lie inside the channel for a mode to be unstable. (ii) An analogue of Howard's semicircle theorem is obtained: if a mode is unstable, then its complex growth rate $\tl \g$ must lie inside a disk of radius $\tl k$. (iii) Since $|\Im \tl \g| < \tl k$ is necessary for a mode to be unstable, it is valuable to find a necessary criterion for a neutrally stable mode to satisfy $\tl \g_i < \tl k$, as such a mode may be viewed as a candidate to develop an instability. We show that such a necessary condition is the lower bound $M \geq (1/2) [1 + 1/ 2\tl k^2]^{1/2}$. When this condition is met, a neutrally stable mode can exist whose critical layer is inside the channel. (iv) Finally, we find a lower bound $|\tl \g_i| > \tl k(1/M - 1)$ for neutrally stable modes. In particular, this implies that level crossings between neutrally stable conjugate modes (with eigenvalues $i \tl \g_i$ and $- i \tl \g_i$) cannot occur for a subsonic background flow ($M<1$). This is significant since we find numerically that instabilities typically begin at a crossing between the lowest mode and its conjugate.

%---------------

\paragraph{Integral identity for stability criteria.} We divide (\ref{w-diff-eqn}) by $\G^2$ and combine the first two terms using the Leibniz rule, to get
	\beq
	\left(\frac{\tl w'}{\G^2}\right)' - \left(\frac{\tl k^2}{\G^2}+M^2 \right) \tl w =0.
	\eeq
Multiplying by $\tl w^*$, this becomes 
	\beq
	\left(\frac{\tl w^* \tl w'}{\G^2}\right)' - \frac{|\tl w'|^2}{\G^2} - \left( \frac{\tl k^2}{\G^2} + M^2 \right)| \tl w|^2 =0.
	\eeq
Integrating in $\tl y$, the first term vanishes due to the BCs $[\tl w' (\pm 1 ) = 0]$ and we get 
	\beq
	 - \int_{-1}^{1}\frac{1}{\G^2}(|\tl w'|^2 + \tl k^2 | \tl w|^2) d\tl y = M^2 \int_{-1}^{1} | \tl w|^2 d \tl y.
	\label{e:integral-id-wstar-times-w-eqn}
	\eeq
Taking the real part, we obtain
	\beq
	 -  \int_{-1}^{1}  \frac{\Re (\G^{*2})}{|\G|^4}(|\tl w'|^2 + \tl k^2 | \tl w|^2) d \tl y = M^2 \int_{-1}^{1} | \tl w|^2 d \tl y.
	\label{e:re-part}
	\eeq
These integral identities will be put to use below in deriving criteria for instability.

%---------------
\subsubsection{Critical layer condition for instability}
\label{s:crit-lay-cond-instab}
%----------------

Noting that $\Im(1/\G^2) = - 2 \tl \g_r (\tl \g_i + \tl k \tl y)/ |\G|^4$ and taking the imaginary part of (\ref{e:integral-id-wstar-times-w-eqn}), we get
	\beq
	 2 \tl \g_r \int_{-1}^{1} (\tl \g_i + \tl k \tl y)\frac{(|\tl w'|^2 + \tl k^2 | \tl w|^2)}{|\G|^4} d \tl y =  0.
	\label{e:im-part}
	\eeq
For an unstable mode ($ \tl \g_r > 0 $), this can happen only if $\tl \g_i + \tl k \tl y$ changes sign in the channel. In other words, there must be a $|\tl y_c| < 1$ such that $\Im \G(\tl y_c) = 0$. Such a layer $\tl y_c = - \tl \g_i/\tl k$ is called a critical layer. Thus, a necessary condition for the mode $\tl \g$ to be unstable is that the critical layer must lie inside the channel. Stated differently, for a mode to be unstable, its wavenumber must be sufficiently large, in the sense that $\tl k > |\tl \g_i|$. We observe that when this condition is met, the CC approximation of \S \ref{const-coeff-lim} breaks down and we enter the level crossing (LC) regime to be discussed in \S \ref{s:stab-trans-vary-M} and \S \ref{s:stab-trans-vary-k}. Pleasantly, our numerical calculations show the onset of instabilities in the LC regime. Moreover, since higher modes have larger $|\tl \g_i|$, it would typically require a higher wave number perturbation to destabilize a higher mode. This is borne out by the numerical results shown in Fig.~\ref{f:gavsk-const-vorticity-la=0}. 

%---------------
\subsubsection{Analogue of semicircle theorem}
%---------------

Now, we establish an analog of Howard's semicircle theorem for isothermal compressible inviscid Couette flow. Denoting $G^2 = (|\tl w'|^2 + \tl k^2 | \tl w|^2)/|\G|^4 > 0$, (\ref{e:re-part}) becomes
	\beq
	 -  \int_{-1}^{1}  (\tl \g_r^2 -(\tl \g_i + \tl k \tl y)^2)G^2 d \tl y=  M^2 \int_{-1}^{1} |\tl w|^2 d \tl y.
	\label{e:re-part-2}
	\eeq
For an unstable mode, $\tl \g_r > 0$ and (\ref{e:im-part}) implies that
	\beq
	\int_{-1}^{1} (\tl \g_i + \tl k \tl y)G^2 d \tl y =  0.
	\label{e:cond-1-sem}
	\eeq
Upon using (\ref{e:cond-1-sem}), (\ref{e:re-part-2}) becomes
	\beq
	 -  (\tl \g_r^2  + \tl \g_i^2) \int_{-1}^{1}G^2 d \tl y +\tl k^2 \int_{-1}^{1} \tl y^2 G^2 d \tl y =  M^2  \int_{-1}^{1}| \tl w|^2 d \tl y.
	 \eeq
Since $|\tl y| < 1$, this implies that
	\beq
	(\tl k^2 -  \tl \g_r^2  - \tl \g_i^2) \int_{-1}^{1}G^2  d \tl y >  M^2  \int_{-1}^{1}| \tl w|^2 d \tl y > 0.
	\label{e:re-part-3}
	\eeq
Since $G^2 \geq 0$, the prefactor on the left must be positive. Thus, for an unstable mode, $\tl \g$ must lie inside a disk of radius $\tl k$ in the complex plane: $\tl \g_r^2 + \tl \g_i^2 < \tl k^2$.

%---------------
\subsubsection{Lower bound on neutrally stable mode eigenvalues}
\label{s:low-bd-neut-eg}
%---------------

If a mode is neutrally stable i.e., $\tl \g_r  = 0$, then $\G^{*2} = - |\G|^2$ and (\ref{e:re-part}) implies
	\beq
	  \int_{-1}^{1}  \frac{|\tl w'|^2}{|\G|^2} d \tl y =  \int_{-1}^{1}\left( M^2 - \frac{\tl k^2}{|\G|^2} \right)| \tl w|^2 d \tl y.
	\label{e:re-part-stable}
	\eeq
The first term is nonnegative, so there must be $|\tl y| < 1$ such that the quantity in parenthesis is nonnegative. Using $\G^2 = -(\tl \g_i + \tl k \tl y)^2$, we get
	\beq
	(\tl \g_i + \tl k \tl y)^2 \geq \tl k^2/M^2 \quad 
	\imply \quad |\tl \g_i| \geq \tl k ( 1/M - 1 ). 
	\label{e:re-part-stable-2}
	\eeq
Thus we have a lower bound on $|\tl \g_i|$ for any neutrally stable mode around subsonic flow. Hence, for $M < 1$, there cannot be a crossing of a neutral mode $\tl \g = i \tl \g_i$ with its conjugate $\tl \g^* = - i \tl \g_i$. As modes are ordered with increasing $\tl \g_i$, this implies a spectral gap between the lowest neutrally stable mode and its conjugate. Interestingly, for $M > 1/2$, (\ref{e:re-part-stable-2}) implies that $|\tl \g_i| > k$ so that $\tl \g$ for the stable mode lies outside  Howard's circle implying that the mode cannot develop an instability for any $k$.

Our numerical results from \S \ref{s:stab-trans-vary-M} and \S \ref{s:stab-trans-vary-k} indicate that the first instability with increasing $k$ or $M$ occurs at a confluence of the lowest lying mode and its conjugate ($\tl \g_0$ and $\tl \g_0^*$). Coupled with this numerical observation, the above spectral gap implies there cannot be any instability for $M < 1$. This is consistent with the results in Figs.~\ref{gavsM-const-vorticity-la=0} and \ref{f:instab-stripe-grnd-mode}.

%--------------------------
\subsubsection{Condition for a neutrally stable mode to have a critical layer}
\label{s:stab-mode-crit-lay}
%---------------------------

Using the differential identities
	\beq
	\frac{\tl w'}{\G} = i \tl k \frac{\tl w}{\G^2} + \left(\frac{\tl w}{\G}\right)' \quad \text{and} \quad \tl w'' = 2i\tl k \left(\frac{\tl w}{\G}\right)'+\left(\frac{\tl w}{\G}\right)'',
	\label{e:w/G} 
	\eeq
the first derivative terms in the $\tl w$ perturbation equation (\ref{w-diff-eqn}) can be eliminated:
	\beq
	\left(\frac{\tl w}{\G}\right)'' - \left(\tl k^2+M^2 \G^2 - 2 \frac{\tl k^2}{\G^2} \right) \frac{\tl w}{\G} =0.
	\label{e:mul-div-by-G}
	\eeq
Multiplying by $(\tl w/\G)^*$ and using the product rule, we get
	\beq
	\left(\frac{\tl w^*}{\G^*}\left(\frac{\tl w}{\G}\right)'\right)' -\left|\left(\frac{\tl w}{\G}\right)'\right|^2- \left(\tl k^2+M^2 \G^2 - 2 \frac{\tl k^2}{\G^2} \right) \left|\frac{\tl w}{\G}\right|^2 =0.
\label{e:mul-by-w-by-G-conj}
	\eeq
Integrating over $ -1 < \tl y < 1$ and using the BCs $(\tl w/\G)'|_{\pm 1} =( -i \tl k\tl w/\G^2)|_{\pm 1}$, we get
	\beq
	- \frac{i \tl k}{\G}\left|\frac{\tl w}{\G}\right|^2 \bigg|^1_{-1}- \int_{-1}^{1}\left|\left(\frac{\tl w}{\G}\right)'\right|^2 d \tl y =  \int_{-1}^{1} \left(\tl k^2+M^2 \G^2 - 2 \frac{\tl k^2}{\G^2} \right)  \left|\frac{\tl w}{\G}\right|^2 d \tl y.
	\label{e:int-out}
	\eeq
Let us now specialize to a neutrally stable mode $(\tl \g_r = 0)$ that has a critical layer, i.e., $|\tl \g_i| < k$. This implies $ (\G/i \tl k)|_{\mp 1}=(\tl \g_i/\tl k ) \mp 1 \lessgtr 0$, so that the LHS is negative. Consequently, 
	\beq
	\tl k^2+M^2 \G^2 - 2 \frac{\tl k^2}{\G^2} < 0
	\quad \text{for some}\quad |\tl y| < 1.
	\label{e:lc-reg-ineq}
	\eeq
Since $|\tl \g_i| < \tl k$, $\G^2 = -(\tl \g_i + \tl k \tl y)^2 > - 4 \tl k^2$. Using this, we deduce a necessary condition for a neutrally stable mode to have a critical layer inside the channel:
	\beq
	M^2 > \frac{1}{4}+\frac{1}{8 \tl k^2}.
	\label{e:necc-inst-cond}
	\eeq
When a neutrally stable mode has a critical layer inside the channel, the critical layer satisfies $\Re \tl \G (\tl y_c) = 0$ in addition to $\Im \G (\tl y_c) = 0$ so that $\tl y_c$ is a regular singularity of (\ref{w-diff-eqn}). Moreover from (\ref{exact-soln}) we find $\tl w'(\tl y_c) = 0$. Thus, the critical layer for such a stable mode is also where the vorticity and density perturbations are extremal. In \S \ref{s:lc-regime} we will relate instabilities to level crossings. It follows that an instability arises (when $\tl k$ or $M$ is varied) when two neighboring local extrema of the vorticity perturbation `annihilate' leaving an unstable mode. On the other hand, for any fixed $\tl k$ if $M < (1/2)(1 + 1/2\tl k^2)^{1/2}$ then $|\tl \g_i| > \tl k$ which from \S \ref{s:crit-lay-cond-instab} implies that all modes are neutrally stable. This is an improvement on the $M < 1/2$ bound of \S \ref{s:low-bd-neut-eg} guaranteeing neutral stability.

\paragraph{Remark.} Eqn. (\ref{e:necc-inst-cond}) allows us to improve on the bound $M < 1$ obtained in \S \ref{s:low-bd-neut-eg} that ensures $|\tl \g_i| > 0$ thereby precluding crossings between conjugate modes. In fact, combining these two results, $M^2 < \max \{1, 1/4 + 1/(8\tl k^2)\}$ is a sufficient condition to avoid such crossings. Thus, for $\tl k < 1/\sqrt{6}$, there is a range of supersonic Mach numbers (see Fig.~\ref{f:instab-stripe-grnd-mode}) for which the Couette flow is neutrally stable. Furthermore, for $\tl k > 1/\sqrt{6}$ there is a range of subsonic Mach numbers for which the neutrally stable modes may admit a critical layer inside the channel but for which resonant interactions between conjugate modes is still forbidden. If such a flow displays an instability, it cannot arise from conjugate mode crossings. Our numerical results in \S \ref{s:lc-regime} indicate that this does not happen since a crossing between the lowest mode and its conjugate precedes any other crossing.

%-------------------------------------------------

%------------------------
\section{Neutrally stable regime: infinite tower of modes}
\label{s:small-k-m-CC}
%------------------------

We find that perturbations to inviscid compressible Couette flow governed by the equations of \S \ref{s:rho-u-v-w-ODE} can manifest a variety of stability characteristics depending on wave and Mach numbers $\tl k$ and $M$. In this section, we examine the stability of this flow by estimating the growth rate $\tl \g$ by solving (\ref{e:v-diff-eq-const-vortex}) and (\ref{w-diff-eqn}) in several limits: small $\tl k$ (\S \ref{s:small-k-regime}), small $M$ (\S \ref{s:small-M-regime}) and large $\Im \tl \g$ (\S \ref{const-coeff-lim}). In each of these limits, we find an infinite tower of neutrally stable modes. Subsequently, in \S \ref{s:lc-regime}, we treat the perturbation equation (\ref{e:v-diff-eq-const-vortex}) via a more general numerical method and discover instabilities arising from level crossings.

%------------------------
\subsection{Small wave number limit}
\label{s:small-k-regime}
%------------------------

We discuss two possibilities for small $\tl k$ which we will interpret as excited and ground modes. 

{\noindent \bf Excited modes $(n \geq 1)$.} First, we suppose that $\tl \g$ has a nonzero (possibly infinite) limit as $\tl k \to 0$ while $\tl v$ and its derivatives have finite limits. In this case, (\ref{e:v-diff-eq-const-vortex}) becomes:
	\beq
	\tl v'' - M^2 \tl \g^2 \tl v = 0.
	\label{e:v-diff-eq-const-vortex-kto0-other-limit}
	\eeq
Viewing $\tl \g \tl v \sim \tl v_t$, we recognize this as the wave equation with propagation speed $\sqrt{1/M}$, which is that of sound only when $M = 1$. Imposing impenetrable BCs, we find a pair of neutrally stable eigenmodes for each positive integer $n$:
	\beq
	\pm \tl \g_n =   \pm \frac{i n \pi}{2M} 	\quad \text{and} \quad
	\tl v_n (\tl y) = e^{\frac{i n \pi \tl y}{2}} + (-1)^{n+1} e^{-\frac{i n \pi \tl y}{2}}.
	\label{e:v-soln-kto0-n}
	\eeq 
Although $\tl v_n (\tl y)$ for both $\tl \g_n$ and $-\tl \g_n$ are the same, the corresponding $\tl \rho_n \propto \tl v_n'$ (\ref{e:vort-rho-pert}, \ref{v-diff-eqn}) and $\tl u_n$ (\ref{u-v-w-P}) have opposite signs. Thus, the eigendisturbances $(\tl \rho, \tl u, \tl v)$ for $\tl \g$ and $- \tl \g$ are linearly independent. 

{\noindent \bf Ground modes $(n=0)$.} Note that in (\ref{e:v-soln-kto0-n}), $n=0$ is not admissible since we assumed that $\tl \g$ has a nonzero limit. However, it turns out that there is a pair of modes for which $\tl \g \to 0$ as $\tl k \to 0$ (see Fig.~\ref{f:im-ga-vs-k-const-vort}) which we turn to now. Let us suppose that $\tl \g = i \tl k s + {\cal O}(\tl k^2)$ where $s$ is a complex constant and $\tl v$ and its derivatives have finite limits as $\tl k \to 0$. Then (\ref{e:v-diff-eq-const-vortex}) becomes
	\beq
\left({\tl k}^2+M^2 \G^2\right) \tl v'' - 2 M^2 \G i \tl k\tl v' -  2 \tl k^2  M^2 \tl v = 0.
\label{e:v-diff-eq-const-vortex-kto0}
	\eeq
To solve this ODE we put $\G / i \tl k \approx s + \tl y$ in the series solution for $\tl v$ (\ref{v-u-expn}) obtained in Appendix \ref{s:power-series-w} to get
	\beq
	\tl v (\tl y) =  i \tl k \left[ \frac{1}{M^2} +  (s + \tl y)^2  + \ldots \right] a_0 +  \left[ 3 \frac{(s + \tl y)}{{i \tl k M^2}} + \tl k \frac{(s + \tl y)^3}{{2 i M^2}} + \frac{i \tl k}{4} (s + \tl y)^5 + \ldots \right] a_3.
	\label{e:full-v-series} 
	\eeq
As $\tl k \to 0$, we retain only the leading terms in each parentheses. The BCs $\tl v(\pm 1) = 0$ become
	\beq
	\colvec{2}{i \tl k( \frac{1}{M^2} +  (s + 1)^2  ) & \frac{3 }{i \tl k M^2} (s + 1)}{i \tl k ( \frac{1}{M^2} +  (s -1)^2  ) & \frac{3}{i \tl k M^2} (s -1)}
	\colvec{2}{a_0}{a_3} = 0.
	\eeq
For a nontrivial solution to exist we must have $s = \pm \sqrt{1+ \frac{1}{M^2}}$ leading to the pair of purely imaginary eigenvalues:
	\beq
	\pm \tl \g_0 = \pm i \tl k \sqrt{1 + 1/{M^2}}.
	\label{e:small-k-ground-mode} 
	\eeq		
It follows that $a_3/a_0 = (\pm 2/3)\tl k^2 M \sqrt{M^2+1}$ is of order $\tl k^2$ which justifies our keeping only the leading terms in (\ref{e:full-v-series}). The corresponding eigenfunctions are quadratic in height $y$:
	\beq
	\tl v (\tl y) =  M^{-2} - 2 s (\tl y + s )  + (\tl y + s )^2.
	\label{e:a3-a0} 
	\eeq
The eigenvalue $\tl \g_0$ in (\ref{e:small-k-ground-mode}) is in quantitative agreement with our numerical results for small $\tl k$ (Fig.~\ref{f:gavsk-const-vorticity-la=0}) as well as with other approximation schemes (Fig.~\ref{f:im-g-vs-k-M=1-la=0-const-vort}). 

Combining (\ref{e:small-k-ground-mode}) with (\ref{e:v-soln-kto0-n}), for small $\tl k$, we have found an infinite tower of neutrally stable modes, $\pm \tl \g_n$ for $n = 0,1,2, \ldots$. These are compressible noninflectional modes that do not have incompressible counterparts. While the eigenvalues of the ground modes $\pm \tl \g_0$ vanish linearly with $\tl k$, those of the excited modes $\pm \tl \g_{n > 0}$ approach nonzero (imaginary) values, as is also visible in Figs.~\ref{f:im-g-vs-k-M=1-la=0-const-vort}, \ref{f:im-g-vs-k-M=9pt5-la=0-const-vort-CC} and \ref{f:gavsk-const-vorticity-la=0}. Interestingly, all these neutrally stable modes are supersonic ($M_p > 1$) for the phase speed of perturbations. Indeed, from (\ref{e:mach-pert}) 
	\beq
	M_p^0 = \sqrt{1 + M^2} \quad \text{and} \quad M_p^{n} = M \tl \g_n/\tl k = n \pi/2 \tl k \quad \text{for} \quad \tl k \ll 1  \quad \text{and} \quad n \geq 1 .
	\eeq

%------------------------
\subsection{Small Mach number limit}
\label{s:small-M-regime}
%------------------------

Assuming $\tl v, \tl v', \tl v''$ and $\tl \g$ all have finite limits as $M \to 0$, the $\tl v$ equation (\ref{e:v-diff-eq-const-vortex}) simplifies to $\tl v'' - \tl k^2 \tl v = 0$ with $\tl \g$ dropping out so that this ceases to be an eigenvalue problem. Moreover, the solutions $A \cosh \tl k y + B \sinh \tl k y$ cannot satisfy the BCs $\tl v(\pm 1) = 0$. Thus, there is no such eigenmode. This may also be seen from the series solution (\ref{v-u-expn}). Indeed, as $M \to 0$, the recursion relation simplifies to $a_{n+4} = \frac{\tl k^2}{(n+4)(n+1)}a_{n+2}$ and the series solution may be summed to give
	\beq
	\tl v(\tl y) = \frac{1}{M^2} \bigg( i \tl k a_0 \cos (\tl \g + i \tl k \tl y) - \frac{3 a_3}{\tl k^2} \sin(\tl \g + i \tl k \tl y) \bigg)
	\label{e:v-Mto0}
	\eeq
Note that $\tl \g$ can be absorbed into the arbitrary coefficients $a_0, a_3$. However, there is no such nontrivial solution that satisfies the BCs $v(\pm 1) = 0$. In fact, the determinant of the coefficient matrix $(3 i/\tl k)\sin(2 i \tl k)$ is independent of $\tl \g$ and nonvanishing for all $\tl k$. Thus, there is no eigenmode with a finite value of $\tl \g$ as $M \to 0$.

%-----------------
\iffalse
As $M \to 0$, the solution for $\tl v$ is given by
	\beq
	\tl v(\tl y) = \frac{(1/M)^2}{i \tl k} \sum_{m=0}^{m = \infty} \frac{-\tl k^{2m+2}}{(2m) !}  \bigg(\frac{\G}{i \tl k}\bigg)^{2m} a_0 +  \frac{(1/M)^2}{i \tl k} \sum_{m=1}^{m = \infty} \frac{\tl k^{2m-2}}{(2m - 1)!}  \bigg(\frac{\G}{i \tl k}\bigg)^{2m-1} 3 a_3 
	\eeq
This can be rewritten as
	\beq
	\tl v(\tl y) = (1/M)^2 \bigg( i \tl k a_0 \cos \G - \frac{3 a_3}{\tl k^2} \sin \G \bigg)
	\label{e:v-Mto0}
	\eeq
Putting the boundary condition $\tl v(\pm 1) = 0$ in equation  (\ref{e:v-Mto0}),
	\beq
	\colvec{2}{i \tl k \cos \G(+1) & -\frac{3}{\tl k^2} \sin \G(+1)}{i \tl k \cos \G(-1) & -\frac{3}{\tl k^2} \sin \G(-1)}
	\colvec{2}{a_0}{a_3} = 0.
	\label{e:bdary-v-Mto0}
	\eeq
For non-triviality of the solution, the determinant has to vanish. The determinant of (\ref{e:det-v-Mto0}) is given by 
	\beq
	\frac{3i}{\tl k} \sin(\G(+1)-\G(-1)) = \frac{3i}{\tl k} \sin(2 i \tl k) . 
\label{e:det-v-Mto0}
	\eeq
Thus, we see the above equation is independent of $\tl \g$, and is always non-zero for any value $\tl k$. Thus, in this limit, there is no eigenvalue as $M \to 0$. The solution (\ref{e:v-Mto0}) satisfies the $\tl v$-equation \ref{e:v-diff-eq-const-vortex} at $M \to 0$, assuming $\tl \g$ is finite at $M to 0$ :  
	\beq
	\tl v''- \tl k^2 \tl v = 0.
	\label{e:v-diff-eq-Mto0-finite-ga}
	\eeq
\fi
%-----------------

A complementary possibility is that $\Im \tl \g$ diverges as $M \to 0$. The simplest way is for $\tl \g = i r/M + r_0 + r_1 M + \cdots$ where $r$ is a complex constant with nonzero real part. Assuming that $\tl v$ and its derivatives have finite limits as $M \to 0$, the $\tl v$ differential equation (\ref{e:v-diff-eq-const-vortex}) becomes:
	\beq
	\tl v''- \left(\tl k^2 - r^2\right) \tl v = 0.
	\label{e:v-diff-eq-Mto0-ga=1/M}
	\eeq
We first consider the possibility that $\tl k^2 - r^2 = 0$ which implies that $\tl \g = \pm i \tl k/M$. In this case (\ref{e:v-diff-eq-Mto0-ga=1/M}) reduces to $\tl v'' (\tl y) = 0$ whose solution upon imposing the BCs is $\tl v(\tl y) \equiv 0$. Interestingly, this leads to a nontrivial disturbance, since from (\ref{u-v-w-P}), $\tl w' = \G M^2 \tl v = 0$ so that the vorticity perturbation $\tl w$ is a constant and from $\tl w = ik \tl v - \tl u'$, $\tl u$ is linear in $\tl y$. We will regard these as the zeroth/ground modes. They are both neutrally stable at small $M$.

On the other hand, assuming $\tl k^2 - r^2 \neq 0$ and imposing impenetrable BCs, we find a pair of eigenmodes for each nonnegative integer $n$:
	\beq
	\tl \g_n = \pm (i/2M) \sqrt{4 \tl k^2 + n^2 \pi^2} 		\quad \text{and} \quad
	\tl v_n (\tl y) = e^{\frac{i n \pi \tl y}{2}} + (-1)^{n+1} e^{\frac{-i n \pi \tl y}{2}}
	\label{e:v-soln-Mto0-1/M}
	\eeq
However, for $n=0$, $\tl \g_0 = \pm  i \tl k/M$ reduce to the ground modes. It is noteworthy that although $\tl v_n (\tl y)$ for $\tl \g_n$ and $-\tl \g_n$ are equal, the corresponding 
	\beq
	\tl \rho_n \to -\frac{\tl \g_n \tl v'}{\tl k^2/M^2 + \tl \g_n^2} 
	= \pm \frac{2 M i}{n^2 \pi^2} \sqrt{4 \tl k^2 + n^2 \pi^2} \: \tl v'
	\eeq
from (\ref{e:vort-rho-pert}) and (\ref{v-diff-eqn}) have opposite signs and make the eigendisturbances $(\tl \rho, \tl u, \tl v)$ for $\tl \g$ and $- \tl \g$ linearly independent.

\begin{figure}[h]
\centering
\begin{subfigure}[]{.3 \textwidth}
\centering
\includegraphics[width=\textwidth]{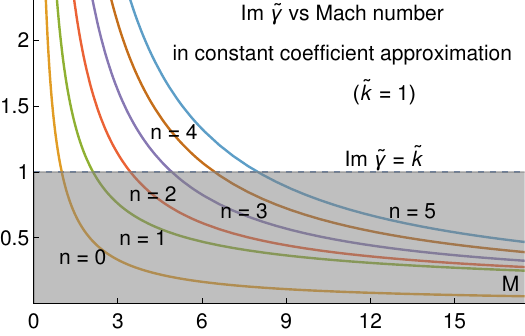}
\caption{\small}
\label{f:gvsM-cclimit-la=0}
\end{subfigure}
\quad
\begin{subfigure}[]{.3 \textwidth}
\centering
\includegraphics[width=\textwidth]{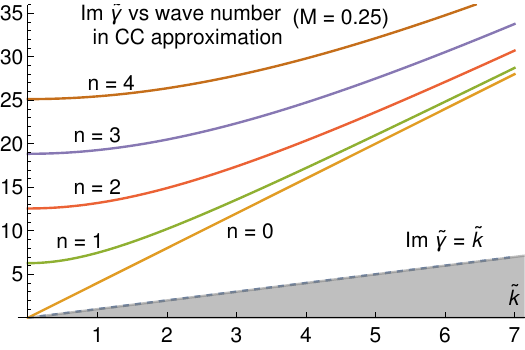} 
\caption{\small}
\label{f:im-g-vs-k-M=1-la=0-const-vort}
\end{subfigure}
\quad
\begin{subfigure}[]{.3 \textwidth}
\centering
\includegraphics[width=\textwidth]{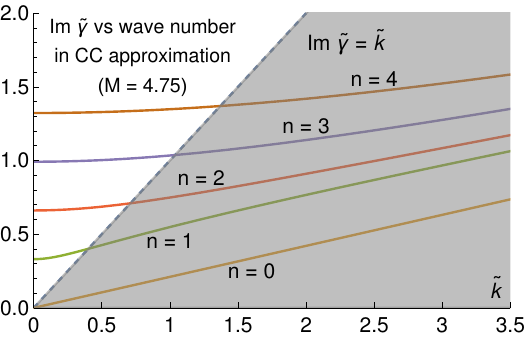} 
\caption{\small}
\label{f:im-g-vs-k-M=9pt5-la=0-const-vort-CC}
\end{subfigure}
\caption{\small (a) $\Im \tl \g_n(M)$ for $\tl k = 1$ and $\Im \tl \g_n( \tl k)$ for (b) $M = 0.25$ and (c) $M = 4.75$ for a few low lying modes in the constant coefficient approximation where $\Re \tl \g_n$ is identically zero. The CC approximation is valid when $|\Im \tl \g| \gg \tl k$ and invalid in the shaded regions. From (a), for any fixed $\tl k$ and $n$, the CC approximation is valid for small $M$. For small enough $M$ (e.g. (b)) the CC approximation is valid for every mode at all values of $\tl k$. For large $\tl k$, $\Im \tl \g_n$ for excited modes asymptote to $\Im \tl \g_0 = \tl k/M$. For larger $M$ (as in (c)) the CC approximation holds only for excited modes and at small $\tl k$. When the CC approximation holds, $\Im \tl \g_n$ form an increasing sequence with no level crossings or instabilities ($\Re \tl \g_n = 0$). }
\label{f:gvsM-cclimit}
\end{figure}

Thus for small $M$ (as for small $\tl k$), we have an infinite sequence of pairs of neutrally stable eigenmodes (labelled by nonnegative integers) with purely imaginary eigenvalues. These are compressible noninflectional modes since all these eigenvalues $\sim 1/M$, and are expelled from the spectrum in the incompressible limit $M \to 0$. This is evident in Fig.~\ref{f:gvsM-cclimit-la=0} and Fig.~\ref{ImgavsM-la=0}. As for small $\tl k$, these modes are supersonic for the phase speed of perturbations (\ref{e:mach-pert}):
	\beq
	M_p^n = \sqrt{1 + n^2 \pi^2/ 4 \tl k^2} \geq 1 \quad \text{for} \quad M \ll 1 \quad \text{and} \quad n \geq 0.
	\eeq
Thus, from (\ref{e:v-diff-eq-Mto0-ga=1/M}), for small $M$, the ground modes $(0,0^*)$ have $M_p = 1$, propagate horizontally and are longitudinal ($\tl u \ne 0$ while $\tl v = 0$) while the higher modes have $M_p > 1$ and display velocity perturbations that are both longitudinal and transverse to the unperturbed flow (see (\ref{e:v-soln-Mto0-1/M})).

A nice feature of this small $M$ limit is that it captures compressional modes which are not present if one takes the strict $M \to 0$ as in (\ref{e:rayleigh-pert}). 

%-------------------
\subsection{High frequency or large \texorpdfstring{$\Im \tl \g$}{Im gamma} limit} 
\label{const-coeff-lim}
%-------------------

Here we turn to another approximation which we refer to as the constant coefficient (CC) limit. It encompasses small $M$ for all modes and small $\tl k$ for excited modes. The idea is to reduce (\ref{w-diff-eqn}) to a constant coefficient equation by supposing that $\G  = \tl \g + i \tl k  \tl y \approx \tl \g$. This happens provided 
	\beq
	|\Im \tl \g| \gg |\tl k |.
	\label{e:cc-approx-valid-ineq}
	\eeq
In particular, in the CC limit, $\tl \g$ cannot be real. In fact, it may be viewed as a limit where eigenmodes display high frequency oscillations. The conditions under which the CC approximation holds will be clarified at the end of this section. 

When $|\Im \tl \g| \gg |\tl k |$, the perturbation equation for $\tl w$ (\ref{w-diff-eqn}) becomes
	\beq
	\tl w''-2 (i  \tl k/\tl \g) \tl w' - ( {\tl k}^2 + M^2 \tl \g^2 ) \tl w = 0.
	\label{w-diff-eqn-const-coeff} 
	\eeq
Putting $\tl w \sim e^{\del \tl y}$, we find that 
	\beq
	\del = \al \pm \beta \quad \text{where} \quad
	\al =  i \tl k /\tl \g \quad 
	\text{and} \quad 
	\beta = \sqrt{\tl k^2 + \tl \g^2 M^2 - \tl k^2 /\tl \g^2},
	\eeq
resulting in $\tl w = A e^{(\al + \beta) \tl y} + B e^{(\al - \beta) \tl y}$. Imposing $\tl w' = 0$ at $\tl y = \pm 1$ we get the pair of equations
	\beq
	\colvec{2}{(\al + \beta) e^{(\al + \beta)} & (\al - \beta) e^{(\al - \beta)}}{(\al + \beta) e^{-(\al + \beta)} & (\al - \beta) e^{-(\al - \beta)}}
	\colvec{2}{A}{B} = 0.
	\eeq
For $A$ and $B$ to not both be zero, the determinant must vanish giving the condition
	\beq
	(\al^2 - \beta^2) \sin(2 i \beta ) = 0.
	\eeq
So either (i) $\al^2 = \beta^2$ or (ii) $2 i \beta  = n \pi$ which implies $\beta^2 + n^2 \pi^2/4 = 0$ where $n \in \mathbb{Z}$. In case (i),
	\beq
	\tl \g^2 + \frac{\tl k^2}{M^2}  = 0 \quad \Rightarrow \quad \tl \g = \pm  \frac{i \tl k}{M}.
	\label{e:n=0-ga}
	\eeq
This eigenvalue is realized provided the CC condition $|\Im \tl \g| \gg |\tl k |$ holds; this happens provided $ M \ll 1$. For $\al = \beta$, we must have $A=0$ while for $\al = - \beta$, $B=0$. In both cases, $\tl w'(\tl y) \equiv 0$ and the corresponding eigenvorticity perturbation must be a constant. These modes turn out to be the ones with smallest $|\Im \tl \g|$ and may be identified with the ground ($n=0$) modes of \S \ref{s:small-k-regime} and \S \ref{s:small-M-regime}. In case (ii), $\tl \g^2$ must satisfy the quadratic equation
	\beq
	-\tl k^2 - \tl \g^2 M^2 + \tl k^2 /\tl \g^2 = n^2 \pi^2/4,
	\eeq
implying that
	\beq
	\tl \g^2 = (1/2M^2) \left[-(\tl k^2 + n^2 \pi^2/4) \pm \sqrt{(\tl k^2 + n^2 \pi^2/4)^2 + 4 \tl k^2 M^2} \right].
	\label{e:gamma-n-const-coeff-prediction}
	\eeq
Since $M^2 > 0$, we get two real and two imaginary eigenvalues $\tl \g$. The real $\tl \g$ are disallowed as they violate the condition $|\Im \tl \g| \gg |\tl k |$. Imposing this requirement on the imaginary eigenvalues $\tl \g = \pm i \sig_n$, we get a condition for the validity of the constant coefficient approximation:
	\beq
	\sig_n = \frac{1}{\sqrt{2} M}\left[(\tl k^2 + n^2 \pi^2/4) + \sqrt{(\tl k^2 + n^2 \pi^2/4)^2 + 4 \tl k^2M^2} \right]^{1/2} 
	\gg |\tl k |.
	\label{eq:ccl}
	\eeq

For fixed values of the parameters $M$ and $\tl k$, this condition is always satisfied for sufficiently large $n^2$. Thus we have an infinite tower of imaginary eigenvalues $\pm \tl \g_n$ for sufficiently large $n \gg 1$. The corresponding vorticity perturbations are
	\beqs
	\tl w_n (y) &=& (\al - \beta) e^{(\al+\beta)\tl y} - (-1)^n(\al + \beta) e^{(\al - \beta)\tl y}  \cr
%	&=& -(i \tl k \Om/\tl \g_n + i n \pi/2L) e^{(-i \tl k \Om/\tl \g_n + i n \pi/2L)\tl y} - (-1)^{n} \; \text{c.c.} \cr
%	&=& (i k U'/\g_n - i n \pi/2L) e^{(i k U'/\g_n + i n \pi/2L)y} -(-1)^n (i k U'/\g_n + i n \pi/2L) e^{(i k U'/\g_n - i n \pi/2L)y} \cr
	&=& 2 i^n e^{-\frac{i \tl k }{\tl \g_n} \tl y}  \left[\frac{\tl k }{\tl \g_n} \sin\left(\frac{n \pi}{2} \left(\tl y-1 \right) \right) - \frac{i n \pi}{2 } \cos \left(\frac{n \pi}{2} \left(\tl y-1 \right) \right) \right].
	\eeqs
Note that this vanishes identically for $n=0$. In fact, when $n = 0$, from (\ref{u-v-w-P}) $\tl v = 1/(\tl \g M^2) \tl w' = 0$ and from the definition of $\tl w$, $\tl u' = 0$ so that $\tl u$ must be a constant. However, a nonzero constant $\tl u$ violates the $\tl w'(\pm 1) = 0$ BC from (\ref{w-and-wpr-intermsof-u}) . Thus, the nontrivial modes are given by $n = 1,2,3, \ldots$. Combining cases (i) and (ii), in the CC approximation we have an infinite sequence of neutrally stable modes $\tl w_n$ with increasing frequencies $\sig_0 < \sig_1 < \sig_2 < \cdots$. The excited modes have supersonic phase speeds $(M^n_p > 1)$ while for the ground mode, $M_p^0 = 1$.

%--------------------

\paragraph*{Validity of CC approximation.} The CC approximation holds provided $|\Im \g| \gg  |\tl k |$ (\ref{e:cc-approx-valid-ineq}). Here, we comment on its validity as the parameters $M$, $\tl k$ and the mode number $n$ are individually varied. 

(i) It is always valid for sufficiently small $M$. It ceases to be reliable for sufficiently large $M$ since $\Im \tl \g \to 0$ as $M \to \infty$. Thus, instabilities may occur for large $M$.

(ii) For excited modes ($n = 1, 2, \ldots$), the CC approximation can be trusted as long as $\tl k$ is sufficiently small. Thus instabilities in these modes can occur only for relatively short wavelength perturbations. The validity of the CC approximation for the lowest mode ($n=0$) is independent of $\tl k$, it is governed by the value of $M$. The same applies to the higher modes when $\tl k$ is sufficiently large. 

(iii) In the CC approximation, $|\Im \tl \g_n| \to n \pi/2 M$ grows with $n$. Thus, for fixed $\tl k$ and $M$, the CC approximation is always valid for large enough $n$.

%------------------------
\section{Level crossing regime: onset of instabilities}
\label{s:lc-regime}
%------------------------

In this section, we numerically solve the linear stability equation (\ref{e:v-diff-eq-const-vortex}) for a horizontal parallel flow with constant background vorticity. We use the methods of Appendix \ref{s:Fredholm} to find the complex growth rates $\tl \g_n$ of various eigenmodes as functions of $M$ and $\tl k$. Where possible, the results are validated using stability theorems and approximations of \S \ref{s:small-k-m-CC}. Interestingly, when we go beyond the regimes of validity of these approximations, we discover new phenomena including level crossings, stability transitions and windows of stable and unstable behavior as $M$ and $\tl k$ are varied.

%-------------------------

As expected from \S \ref{s:small-k-m-CC}, our numerical investigations of (\ref{e:v-diff-eq-const-vortex}) reveal an infinite tower of modes that may be labelled by a nonnegative integer $n = 0,1,2, \cdots$. These modes are neutrally stable in the CC approximation (i.e., $\tl \g_n = i \sig_n$ with $0 < \sig_0 < \sig_1 < \sig_2 \ldots$). Moreover, due to the symmetries of \S \ref{s:dissym}, each such eigenmode has a `partner' mode with eigenvalue $\tl \g_n^* = - i \sig_n$. However, once parameters leave the regime of validity of the CC approximation, we find that modes may start developing instabilities in a sequential manner beginning with the lowest lying mode $n = 0$. Interestingly, the onset of each instability is associated with a level crossing between a pair of adjacent eigenvalues on the imaginary $\tl \g$-axis, as either $M$ or $\tl k$ is varied. 

%------------

%---------------
\subsection{Stability transitions as the Mach number increases}
\label{s:stab-trans-vary-M}
%---------------

\begin{figure}[!h]
\centering
\begin{subfigure}[]{7.5cm}
\centering
\includegraphics[width=7.5cm]{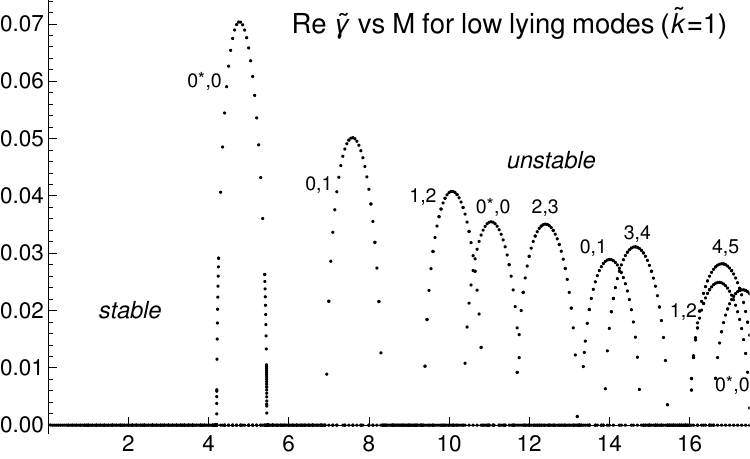} 
\caption{\small}
\label{RegavsM-la=0}
\end{subfigure}
\quad \quad
\begin{subfigure}[]{7.5cm}
\centering
\includegraphics[width=7.5cm]{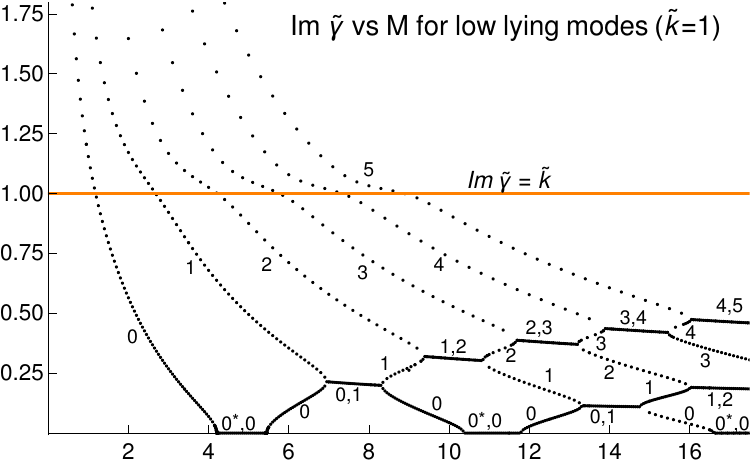}
\caption{\small}
\label{ImgavsM-la=0}
\end{subfigure}
\caption{\small (a) $\Re \tl \g$ and (b) $\Im \tl \g$ vs $M$ for $\tl k = 1$. The flow is stable for small $M$ and unstable modes are manifested as `arches' in (a). As $M$ grows, an increasing number of low-lying modes undergo stability transitions which occur at level crossings, many of which are visible in (b). The number of modes that participate in the instability is expected to grow with $M$ although the severity of the instabilities decrease with $M$ as shown in (a). In (b) when levels with equal $\Im \tl \g$ (e.g. $(0,1)$) split, the one with larger $|\Im \tl \g|$ is assigned the larger mode number. Note that we only show the modes with $\Re \tl \g, \Im \tl \g \geq 0$. For each such $\tl \g$, there are up to 3 other modes related to it via the symmetries of \S \ref{s:dissym}.}
	\label{gavsM-const-vorticity-la=0}
\end{figure}

To illustrate the pattern of these instabilities, we imagine increasing $M$ from $0$ holding $\tl k$ fixed. As $M \to 0$, all $\tl \g_n \to \pm i \infty$ in the manner expected from \S \ref{s:small-M-regime} (see Fig.~\ref{f:gvsM-cclimit-la=0}) and the flow is neutrally stable (for all $\tl k$). In other words, each $\tl \g_n$ is imaginary and decreases in magnitude with increasing $M$ as in Fig.~\ref{ImgavsM-la=0}. As $M$ increases, each eigenmode (beginning with $n = 0$) eventually leaves the CC regime [the condition $|\Im \tl \g | \gg |\tl k|$ (\ref{e:cc-approx-valid-ineq}) is violated] and subsequently enters a level-crossing regime ($|\Im \tl \g | < |\tl k|$) where it goes on to develop an instability ($\Re \tl \g \ne 0$) as manifested in the arches of Fig.~\ref{RegavsM-la=0}. The first instability occurs when $\tl \g_0$ vanishes at a finite $M$ ($\approx 4.203$ for $\tl k = 1$) where it merges with its partner mode $- \tl \g_0$. As $M$ is further increased, they split into a pair of real eigenvalues $-\tl \g_{0^*,0} < 0 < \tl \g_{0^*,0}$ signaling an instability. However, the instability persists only for a small interval in $M$, by the end of which the real eigenvalues merge once again and split into the imaginary pair $\tl \g_0, - \tl \g_0$. The next instability involves a similar merger-demerger encounter between the modes $n =0$ and $n = 1$.

%---------------

%---------------

In fact, we find a common pattern to stability-instability transitions that we now describe. To begin with, each such transition is a local phenomenon involving an adjacent pair of modes (the remaining modes are spectators and may be stable or unstable during this transition). (i) Each stable to unstable transition is associated with the confluence of two imaginary eigenvalues (e.g., $\tl \g_1, \tl \g_2$ or $\tl \g_0, \tl \g_0^* = - \tl \g_0$) on the imaginary $\tl \g$ axis and is manifested by the opening of an arch in the graph of $\Re \tl \g$ in Fig.~\ref{RegavsM-la=0}. (ii) Within the unstable interval of $M$ (arch in $\Re \tl \g$), the eigenvalues that had met on the imaginary axis split off into the pair $\tl \g, - \tl \g^*$. Their real parts increase in magnitude to a maximal value and then decrease to zero. (iii) At this point the instability arch closes signaling an unstable to stable transition associated with the meeting of two complex eigenvalues $\tl \g, - \tl \g^*$ on the imaginary $\tl \g$ axis. (iv) In each window of neutral stability, the eigenvalues that had met on the imaginary axis move apart on the imaginary axis till each of them meets another imaginary eigenvalue at which point a stable to unstable transition occurs as in (i).  The cycle (i)--(iv) repeats itself as $M$ is increased with an increasing number of modes participating in instabilities. Accounting for the tower of eigenvalues, this results in a sequence of $M$ intervals where a pair of adjacent modes $n, n+1$ participate in an instability arch. It is noteworthy that the instability arches may overlap leading to multi-mode instabilities. In fact, every mode eventually develops an instability at some $M$ leading to an infinite sequence of instability arches, though the strength of the instability tends to decrease with $M$. More precisely, when Fig.~\ref{ImgavsM-la=0} is viewed as a directed graph with increasing $M$, the height of instability arches (in Fig.~\ref{RegavsM-la=0}) is a decreasing function along any edge path.

%----------------------
\subsection{Stability transitions as the wavenumber increases}
\label{s:stab-trans-vary-k}
%----------------------

\begin{figure}[!h]
\centering
\begin{subfigure}[]{7.5cm}
\centering
\includegraphics[width=7.5cm]{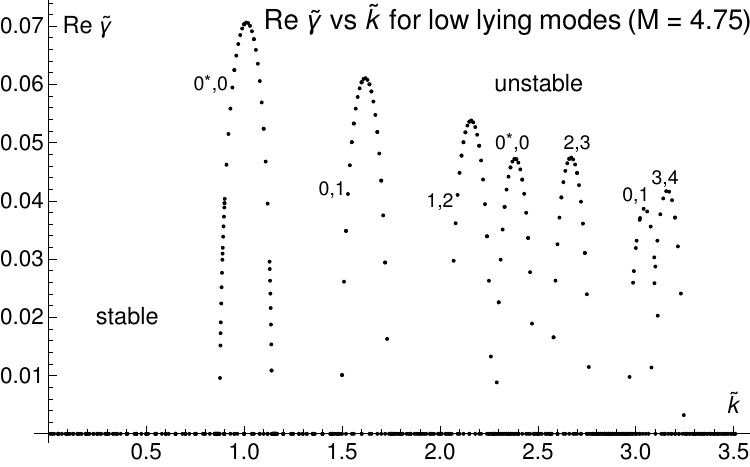} 
\caption{\small}
\label{f:re-ga-vs-k-const-vort}
\end{subfigure}
\quad \quad
\begin{subfigure}[]{7.5cm}
\centering
\includegraphics[width=7.5cm]{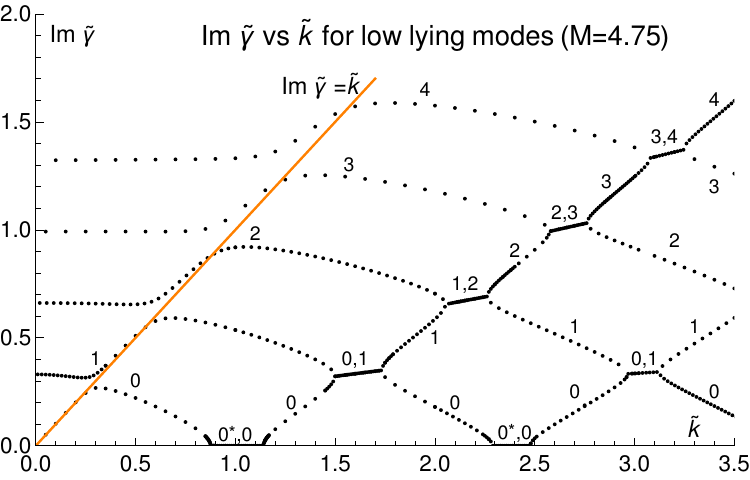}
\caption{\small}
\label{f:im-ga-vs-k-const-vort}
\end{subfigure}
\caption{\small (a) $\Re \tl \g$ and (b) $\Im \tl \g$ vs wave number $\tl k$ for $M = 4.75$. (a) shows that the flow is stable for small $\tl k$. As $\tl k$ grows an increasing number of modes undergo stability transitions which occur at level crossings. The number of modes participating in the instabilities grows with $\tl k$ although the severity of the instabilities decrease with $\tl k$ as happened with increasing $M$ in Fig.~\ref{gavsM-const-vorticity-la=0}. 
}
	\label{f:gavsk-const-vorticity-la=0}
\end{figure}

As Fig.~\ref{f:gavsk-const-vorticity-la=0} indicates, the pattern of instabilities that develop as $\tl k$ is varied holding $M$ fixed is similar to the pattern described above, where $M$ was varied holding $\tl k$ fixed.  We distinguish between two regimes: (a) small $M < 1$ and (b) moderate $M > 1$. (a) For small enough $M$, all modes are neutrally stable to perturbations of any wavenumber, as shown in \S \ref{s:small-M-regime} and in Fig.~\ref{f:im-g-vs-k-M=1-la=0-const-vort}. (b) For larger $M$, the behavior for small $\tl k$ is captured by our small $\tl k$ expansion of \S \ref{s:small-k-regime} with all modes being neutrally stable. In particular, as $\tl k \to 0$, from (\ref{e:v-soln-kto0-n}) and (\ref{e:small-k-ground-mode}), all the eigenvalues have finite limits
	\beq
	\tl \g_0 \sim i \tl k \sqrt{1 + 1/M^2} \quad \text{and} \quad 
	\tl \g_{n \geq 1} \to i n \pi/2M. 
	\eeq
However as $\tl k$ grows, the small $\tl k$ and CC approximations fail and every mode eventually becomes unstable, as illustrated in Fig.~\ref{f:gavsk-const-vorticity-la=0}. These instabilities arise at level crossings between adjacent modes. The first instability occurs when the two imaginary eigenvalues $\tl \g_0$ and $-\tl \g_0$ merge and split into a pair of real eigenvalues $-\tl \g_{0^*,0} < 0 < \tl \g_{0^*,0}$ which exist for a small interval of wavenumbers. This unstable window leads to the first arch in Fig.~\ref{f:re-ga-vs-k-const-vort} which ends when the real eigenvalues coalesce into the imaginary pair $\tl \g_0, - \tl \g_0$. The succeeding instabilities involve similar merger-demerger encounters between adjacent modes ($0,1$ followed by $1,2$ and $0,0$ etc.). Thus, there is a sequence of wavenumber intervals where a pair of adjacent modes $n, n+1$ participate in an instability arch. Although there appear to be an infinite number of instability arches, the strength of the instabilities (heights of arches) tend to decrease with $\tl k$ as happened with increasing $M$. However, unlike in the latter case, as $\tl k$ grows there is no bound on the magnitude of $\Im \tl \g$ at which level crossings occur; this is expected from the critical layer condition of \S \ref{s:crit-lay-cond-instab}.

%---------------
\subsection{Instability stripes in the \texorpdfstring{$\tl k$-$M$}{k-M} plane} 
\label{s:instripes}
%---------------

\begin{figure}[!h]
\centering
\includegraphics[width=7.5cm]{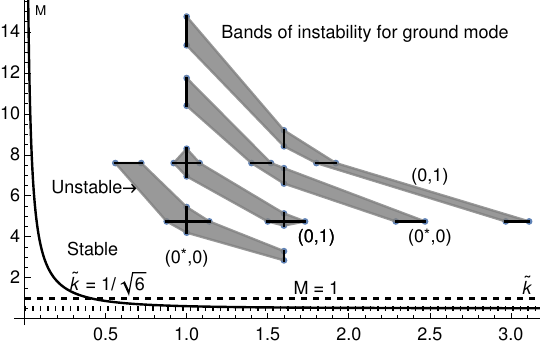} 
\caption{\small Solid line segments are windows of instability in the lowest lying mode ($n=0$) for indicated values of $\tl k$ and $M$. Combining these, shaded regions roughly indicate unstable regions in the $\tl k$-$M$ plane for the ground mode $n=0$. Although not indicated, these regions extend to larger values of $\tl k$ and $M$. The pattern of alternating bands of neutrally stable and unstable behavior repeats as we go further out in the first quadrant of the $\tl k$-$M$ plane. The successive instability stripes alternatively involve a merger of the $(0^*,0)$ and $(0,1)$ modes. Note from \S \ref{s:low-bd-neut-eg} that for $M < 1/2$ (dotted horizontal line) there are no instabilities for any $\tl k$. On the other hand, no such lower bound on $\tl k$ would guarantee neutral stability for all $M$: given a $\tl k$ one expects an instability provided $M$ is large enough. The Remark in \S \ref{s:stab-mode-crit-lay} shows that below the solid curve (for $\tl k < 1/\sqrt{6}$) and the dashed horizontal line (for $\tl k > 1/\sqrt{6}$), there cannot be any instabilities.}
	\label{f:instab-stripe-grnd-mode}
\end{figure}

For a given mode $n$, it is interesting to find the regions of instability in the $\tl k$-$M$ parameter plane. Unlike in the viscous case, there are no curves of marginal stability. Instead, we have regions of instability ($\Re \tl \g_n \ne 0$) and regions of marginal stability ($\Re \tl \g_n = 0$). Here, we find these regions for the ground mode $n = 0$. To this end, we numerically determine the first 4 instability arches for the ground mode with increasing $M$ holding $\tl k=1$ and $\tl k = 1.6$ fixed. The same is done with increasing $\tl k$ keeping $M=4.75$ and $M =7.6$ fixed. As shown in Fig.~\ref{gavsM-const-vorticity-la=0} and \ref{f:gavsk-const-vorticity-la=0}, there are successive windows of instability associated with level crossings of the $n=0$ mode with either its conjugate mode $n = 0^*$ or with the $1^{\rm st}$ excited mode $n=1$. Combining these, we find that the $\tl k$-$M$ plane is partitioned into an infinite sequence of alternating stripes of neutrally stable and unstable behavior. This is indicated in Fig.~\ref{f:instab-stripe-grnd-mode} with unstable (grey) stripes interspersed with neutrally stable (white) regions. Unstable bands associated to $(0^*,0)$ mode mergers alternate with those associated to $(0,1)$ mergers. Amusingly, a similar pattern of stripes separates regions where a coin toss results in a heads or tails in the (initial) excess height $-$ rotational energy plane \cite{vlad-vul-ric}. 

We may infer from Figs.~\ref{gavsM-const-vorticity-la=0} and \ref{f:gavsk-const-vorticity-la=0}, that a similar zebra like pattern is expected for higher modes as well, with instability bands alternating between $(n-1, n)$ and $(n, n+1)$ mode mergers. As the mode number grows, the entire pattern of instability stripes shifts towards the North-East. In fact, the saturation of inequality (\ref{eq:ccl}) (for any given $n \geq 1$) should provide a rough boundary beyond which bands of instability should form. The instability stripes for distinct modes may of course overlap, just as with the instability arches in Figs.~\ref{RegavsM-la=0} and ~\ref{f:re-ga-vs-k-const-vort}. Although the neutrally stable region is expected to shrink upon including unstable regions from additional modes, the strength of the instabilities declines as both $M$ and $\tl k$ grow.

\paragraph{Remark.} Our results suggest that the first instability with increasing $M$ or $\tl k$ is the one arising from a resonant interaction between the ground modes [$(0^*,0)$ merger]. It would be reassuring to prove that a level crossing between any other pair of modes (for which $|\tl \g_i| < \tl k$) cannot precede the $(0^*,0)$ merger. Note that for any fixed $M$ and $\tl k$, our results from \S \ref{const-coeff-lim} and \S \ref{s:crit-lay-cond-instab} imply that sufficiently excited modes cannot be involved in level crossings and are guaranteed to be neutrally stable.

%------------------------
\subsection{Canonical power law form for growth rate near level crossings}
\label{s:canon-form-eval-level-cross}
%------------------------

For fixed $\tl k$, consider a neutrally stable to unstable transition at $M = M_c$ at which a pair of modes [say $n=0$ and its symmetric counterpart $n' = 0^*$ or ($n$ and $n' = n \pm 1$)] undergo a level crossing (left end of an instability arch in Fig.~\ref{RegavsM-la=0}). Thus, for $M < M_c$, $\tl \g_0$ and $\tl \g_0^*$ (or $\tl \g_n$ and $\tl \g_{n'}$) are both imaginary and for $M > M_c$ they acquire nonzero real parts. Moreover, let us use the symbol $\tl \g_c$ for $\tl \g(M_c)$ which is always imaginary. Here, we propose a canonical form for the growth rates $\tl \g$ of these neighboring modes near the level crossing. Fig.~\ref{gavsM-const-vorticity-la=0} indicates that both $\Re \tl \g$ and $\Im \tl \g$ display power-law behaviors as functions of $|M - M_c|$. In fact, if we view the eigenvalues $\tl \g$ as solutions of the characteristic equation $\det(Q -\tl \g I) = 0$ for the differential operator in (\ref{e:compr-cons-Orr-Somm}), then in the vicinity of $M_c$, it is reasonable to suppose that only the $2 \times 2$ block associated to the modes participating in this crossing are relevant while the contributions of other modes may be ignored. Accounting for the requirements that (i) for $M < M_c$ the roots are purely imaginary, (ii) when $M = M_c$, the roots merge at $\tl \g_c$ and (iii) for $M > M_c$ the roots take the form $\tl \g, - \tl \g^*$ dictated by symmetry A of \S \ref{s:dissym}, we propose the quadratic characteristic equation
	\beq
	(\tl \g - \tl \g_c)^2 + A^2 (M_c - M)=0 \quad \text{for small} \quad |M - M_c|
	\label{e:quad-eqn-g-near-Mc}
	\eeq	
at the left end of an instability arch. Here $A$ is a real parameter that could depend on $\tl k$ and the modes involved in the transition. Thus, for small $|M - M_c|$, the eigenvalues are the branches of the 2-valued function $\tl \g(M)$ defined by (\ref{e:quad-eqn-g-near-Mc}) and are given by
	\beq
	\tl \g (M) = \tl \g_c \pm i A \sqrt{M_c - M}.
	\label{e:g-sqrt-0thorder-pm}
	\eeq
This leads to the asymptotic behaviors
	\beqs
	\Im  (\tl \g - \tl \g_c) \approx A \sqrt{M_c - M}  \quad \text{for} \quad M < M_c \cr
	\Re  (\tl \g - \tl \g_c)  \approx A \sqrt{M - M_c}  \quad \text{for} \quad M > M_c.
	\label{e:sqrt-behav-re-im-g}
	\eeqs
According to (\ref{e:g-sqrt-0thorder-pm}) $\Im \tl \g$ is constant (and equal to $- i \tl \g_c$) for $M > M_c$. However, Fig.~\ref{ImgavsM-la=0} indicates a linear dependence on $M$. This may be captured by adding a purely imaginary linear correction term to (\ref{e:g-sqrt-0thorder-pm}). Thus for $|M - M_c| \ll 1$ we have:
	\beq
	\tl \g \approx \tl \g_c \pm i A \sqrt{M_c - M} + iB (M_c-M).
	\eeq
The corresponding equations near an unstable to neutrally stable transition at the right end of an instability arch takes the same form as above except that $M-M_c$ is replaced with $M_c-M$. 

We may extend the canonical form (\ref{e:quad-eqn-g-near-Mc}) to include the counterparts $\tl \g^*, - \tl \g$ of the above eigenvalues $\tl \g, -\tl \g^*$ obtained via symmetry B of \S \ref{s:dissym}. These four eigenvalues in the vicinity of a neutrally stable to unstable transition at $M_c$ may be modeled as the roots of the following biquadratic equation with real coefficients
	\beq
	(\tl \g^2 -  \tl \g_c^2 + A^2(M_c - M))^2 + 4 \tl \g_c^2 A^2(M_c - M)= 0.
	\eeq

\begin{figure}[!h]
\centering
\begin{subfigure}[]{7.5cm}
\centering
\includegraphics[width=7.5cm]{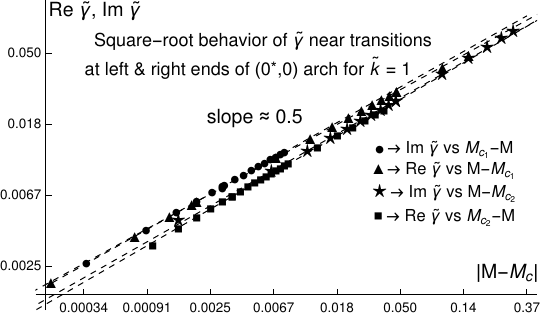} 
\caption{\small}
\label{f:log-ga-vs-log-M-00-k-1}
\end{subfigure}
\quad \quad
\begin{subfigure}[]{7.5cm}
\centering
\includegraphics[width=7.5cm]{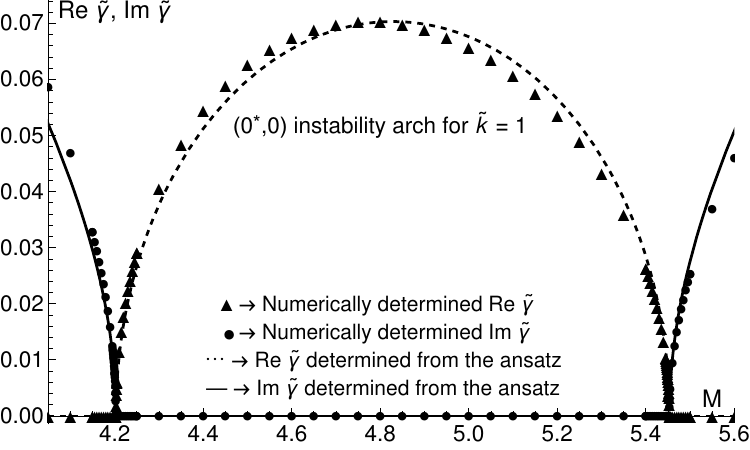}
\caption{\small}
\label{f:ga-global-arch-n00}
\end{subfigure}
\caption{\small (a) Log-Log plots of $\Re \tl \g$ and $\Im \tl \g$ vs $|M - M_c|$ for the stable to unstable and unstable to stable transitions at either end ($M_{c_1} = 4.203, M_{c_2} = 5.4535$) of the $(0^*,0)$ instability arch of Fig.~\ref{gavsM-const-vorticity-la=0}. Here $\tl \g_c = 0$ and $\tl k = 1$. The fitted slopes $0.50$ ($\Im \tl \g$) and $0.49$ ($\Re \tl \g$) at the left end of the arch ($ M_{c_1}$) as well as $0.51$ ($\Im \tl \g$) and $0.52$ ($\Re \tl \g$) at the right end of the arch ($ M_{c_2}$) confirm the square-root power law behavior proposed in (\ref{e:sqrt-behav-re-im-g}). (b) Comparison of model (\ref{e:global-arch}) with numerical data around this arch.}
	\label{f:instability-arch-n00-k=1}
\end{figure}

To test the canonical form (\ref{e:sqrt-behav-re-im-g}), we fit our numerically determined $\tl \g$ at either end of the first instability arch $(0^*,0)$. The log-log plots of Fig.~\ref{f:log-ga-vs-log-M-00-k-1} show that both $\Re \tl \g$ and $\Im \tl \g$ display square-root power law behaviors as functions of $|M - M_c|$ at either end of the arch. This confirms our hypothesis that for the stability transition at $M = M_c$, the characteristic equation for $Q$ admits  the quadratic factor in (\ref{e:quad-eqn-g-near-Mc}) associated to the two crossing modes.

\begin{figure}[!h]
\centering
\begin{subfigure}[]{7.5cm}
\centering
\includegraphics[width=7.5cm]{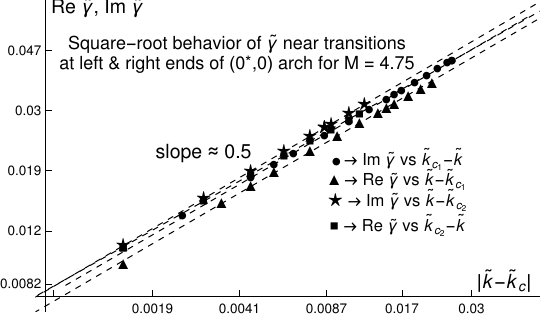} 
\caption{\small}
\label{f:log-ga-vs-log-k-00-M-4pt75}
\end{subfigure}
\quad \quad
\begin{subfigure}[]{7.5cm}
\centering
\includegraphics[width=7.5cm]{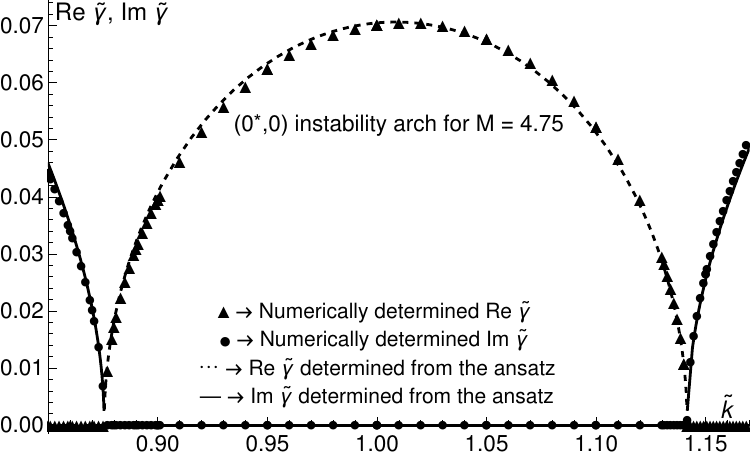}
\caption{\small}
\label{f:ga-global-arch-n00-k-dep}
\end{subfigure}
\caption{\small (a) Log-log plots of $\Re \tl \g$ and $\Im \tl \g$ vs $|\tl k - \tl k_c|$ for the stable to unstable and unstable to stable transitions at either end ($\tl k_{c_1} = 0.8755, \tl k_{c_2} = 1.1415 $) of the $(0^*,0)$ instability arch of Fig.~\ref{f:gavsk-const-vorticity-la=0}. Here $\tl \g_c = 0$ and $M = 4.75$. The fitted slopes $0.50$ ($\Im \tl \g$) and $0.51$ ($\Re \tl \g$) at the left end of the arch ($ \tl k_{c_1}$) as well as $0.51$ ($\Im \tl \g$) and $0.49$ ($\Re \tl \g$) at the right end of the arch ($ \tl k_{c_2}$) confirm the square-root power law behavior. (b) Comparison of model for $\tl \g(\tl k)$ (\ref{e:global-arch}) with numerical data around first $(0^*,0)$ instability arch for $M=4.75$. The value of $\log A = -1.29$ from the analogue of (\ref{e:intercept-A-predic-by-model}) roughly matches the numerical intercepts ($ \log A \approx -1.28, -1.33, -1.20$ and $-1.32$) in (a).}
	\label{f:instability-arch-n00-M=4pt75}
\end{figure}

\paragraph*{Ansatz for an instability arch.} We may combine the behaviors at both ends $M_{c_1} < M_{c_2}$ of an instability arch with the linear behavior of $\Im \tl \g$ to propose a simple interpolating functional form for $\tl \g$ in the neighborhood of the arch (i.e., $M_{c_1} - \eps \lesssim M \lesssim M_{c_2} + \eps$ for small $\eps$) \small
	\beq
	\tl \g = \bigg(\frac{\tl \g_{c_1}+\tl \g_{c_2}}{2}\bigg) 
	+ \left[\frac{\tl \g_{c_1}-\tl \g_{c_2}}{M_{c_1}-M_{c_2}} \right]
	\left[ M-\frac{M_{c_1}+M_{c_2}}{2} \right]  
	\pm \frac{ 2 i  \Re{\tl \g_{\rm mid}}}{M_{c_2}-M_{c_1}} \sqrt{(M_{c_1} - M)(M_{c_2} - M)}.
	\label{e:global-arch}
	\eeq \normalsize
Here $\tl \g_c = \tl \g(M_c)$ and $\tl \g_{\rm mid} = \tl \g \left(\frac{M_{c_1}+M_{c_2}}{2}\right)$. Fig.~\ref{f:ga-global-arch-n00} shows that this form roughly captures the behavior of $\tl \g$ over the whole level crossing merger-demerger of a pair of neighboring eigenmodes. Moreover, (\ref{e:global-arch}) predicts a value for the coefficient in (\ref{e:sqrt-behav-re-im-g}): 
	\beq
	A = 2 \Re\tl \g_{\rm mid}/\sqrt{M_{c_2}-M_{c_1}}.
	\label{e:intercept-A-predic-by-model}
	\eeq
For the first $(0^*,0)$ arch, this suggests $\log A \approx -2.07$ which roughly agrees with the intercepts ($\log A = -1.97, -2.04, -2.11, -2.10$) from the straight line fits in Fig.~\ref{f:log-ga-vs-log-M-00-k-1}.

An entirely analogous canonical form for $\tl \g$ holds near stability transitions involving neighboring modes when $\tl k$ is varied holding $M$ fixed. The $|\sqrt{k - k_c}|$ behaviors of $\Re (\tl \g-\tl \g_c)$ and $\Im(\tl \g - \tl \g_c)$ are demonstrated in Fig.~\ref{f:instability-arch-n00-M=4pt75}.

%------------------------
\section{Continuous spectrum of perturbations to Couette flow}
\label{s:cont-spect}
%------------------------

It turns out that the eigenvalue problem for compressible Couette flow (\ref{cons-pert}) admits a continuous spectrum of imaginary growth rates that we deduce by solving the $\tl w$ equation (\ref{w-diff-eqn}). However, these correspond to $\tl w$ perturbations that fail to be smooth at a critical layer. The approach we follow is a compressible analogue of the method adopted by Drazin in Chapt. 8 of \cite{drazin-book} for incompressible Couette flow. Alternatively one may use the branch cut discontinuities of the Resolvent/Green's functions to examine the continuous spectrum, as done by Case \cite{CaseKM} for incompressible Couette flow. In either approach, one chooses to work with one of the second order ODES (Case \& Drazin work with $\tl v$ while we choose to work with $\tl w$). We will comment on other possibilities at the end of this section.

For definiteness, we consider the ODE (\ref{w-diff-eqn}) for $\tl w$. We will define the critical layer $\tl y = \tl y_{\tl \g}$ by the condition that the Doppler-shifted growth rate $\G(\tl y_{\tl \g})$ vanishes (this reduces to the definition in \S \ref{s:stab-thrm} since $\tl \g$ will be imaginary). For those $\tl \g$ for which the critical layer lies inside the vertical channel $|\tl y| < 1$, (\ref{w-diff-eqn}) admits a patched solution with $\tl v$ being continuous across the critical layer and obeying impenetrable outer BCs. It turns out that such a critical layer exists for any $\tl \g \in \{i \tl k  \tl y| -1 < \tl y < 1\}$. For any such $\tl \g$, the critical layer is given by $ \tl y_{\tl \g} = i \tl \g/\tl k$. Thus, in addition to the discrete spectrum discussed in \S \ref{s:lc-regime}, there is a continuous spectrum along the imaginary axis consisting of $-i \tl k  < \tl \g < i \tl k $. Although the corresponding eigenfunctions depend on both $M$ and $\tl k$, the range of this continuous spectrum only depends on $\tl k$. Interestingly, the continuous spectrum overlaps with a discrete eigenvalue when the latter lies in the level crossing regime ($|\Im \tl \g| < \tl k$). However, it is noteworthy that the wave functions corresponding to these discrete and continuous spectra belong to different regularity classes (smooth and $C^2$ respectively).

We begin our analysis of the continuous spectrum by observing that at the critical layer where $\G(\tl y_{\tl \g}) = 0$, (\ref{w-diff-eqn}) reduces to $\tl w' (\tl y) = 0$. Thus, it is natural to divide the channel into two layers above and below the critical layer $\tl y = \tl y_{\tl \g}$ and seek a patched solution with (i) impenetrable outer BCs $\tl w'(\pm 1) = 0$ and (ii) continuity of $\tl v$ across the critical layer. We note that any such patched solution of (\ref{w-diff-eqn}) would automatically satisfy $\tl w' = 0$ when the critical layer is approached from either side. It is also noteworthy (from (\ref{u-v-w-P})) that $\tl v = 0$ is equivalent to $\tl w' = 0$ as long as $\G \ne 0$ as happens at outer boundaries. This equivalence fails at the critical layer.

To construct such a patched solution, we observe that $\tl w$ in each of the layers can be written in terms of the series solutions of (\ref{w-diff-eqn}) introduced in (\ref{exact-soln}) of Appendix \ref{s:power-series-w}, which may be expressed as
	\beq 
	\tl w(y) = a_0 \: E(\G(\tl y)) + a_3 \: O(\G(\tl y)),
	\eeq
where $ E(-\G(\tl y)) = E(\G(\tl y))$ and $ O(-\G(\tl y)) = -O(\G(\tl y))$ are even and odd functions of $\G$. Explicitly, from (\ref{exact-soln}), 
	\beq 
	E(\G) = 1 - \frac{\tl k^2}{2}  \bigg(\frac{\G}{i \tl k }\bigg)^2 - \tl k^2 \frac{(\tl k^2  + 2 M^2 )}{8 }\bigg(\frac{\G}{i \tl k }\bigg)^4- \tl k^4 \frac{(\tl k^2 - 2 M^2 )}{144 } \bigg(\frac{\G}{i \tl k }\bigg)^6 + \ldots
	\eeq  
and 
	\beq
	O(\G) = \bigg(\frac{\G}{i \tl k }\bigg)^3 + \frac{\tl k^2}{10} \bigg(\frac{\G}{i \tl k}\bigg)^5 + \tl k^2 \frac{(\tl k^2  - 10 M^2)}{280} \bigg(\frac{\G}{i \tl k }\bigg)^7 + \ldots.
	\eeq
So
	\beq 
	\tl w'(\tl y) = i \tl k  [ a_0 E'(\G) + a_3 O'(\G) ]. 
	\eeq
We see from these expansions that $\tl w'(\tl y)$ vanishes on the critical layer as anticipated. Let $\tl w_{+}$ and $\tl w_{-}$ be the solutions above ($1> \tl y > \tl y_{\tl \g}$) and below ($-1 < \tl y < \tl y_{\tl \g}$) the critical layer respectively. The outer BCs $\tl w'(\pm 1) = 0$ require that
	\beq  
	E'(\G(\pm 1)) a_0^\pm + O'(\G(\pm 1)) a_3^\pm = 0,
	\eeq
where $ \G(\pm 1) = \tl \g \pm i \tl k  = - i \tl k ( \tl y_{\tl \g} \mp 1 )$. It is reasonable to suppose that for generic values of parameters ($\tl k, M$) and any $\tl \g \in (- i \tl k , i \tl k )$, $E'(\G(\pm 1))$ and $O'(\G(\pm 1))$ are not identically zero for all such $\tl \g$. Thus, we may solve for $a_3^\pm$ in terms of $a_0^\pm$:
	\beq
 a_3^\pm = - \frac{E'(\G(\pm 1))}{O'(\G(\pm 1))} a_0^\pm 	=  \frac{-\tl k^2 \left( \frac{\G (\pm 1)}{i \tl k} \right) - \tl k^2 \frac{(\tl k^2  + 2 M^2)}{2 } \left(\frac{\G (\pm 1)}{i \tl k } \right)^3-\ldots}{3 \left( \frac{\G (\pm 1)}{i \tl k } \right)^2 + \frac{\tl k^2}{2}  \left(\frac{\G (\pm 1)}{i \tl k }\right)^4 +\ldots}  a_0^\pm
	\label{a-3-1-ext-bound}
	\eeq
Now, we are left with 2 undetermined coefficients $a_0^\pm$. To relate these, we require that $\tl v$ be continuous across the critical layer:   
	\beq
	\tl v_+ (\tl y_{\tl \g}) = \tl v_- (\tl y_{\tl \g}) 
	\quad \imply \quad 
	a_2^+ = a_2^- \quad
	\imply \quad a_0^+ = a_0^-
	\eeq
using (\ref{v-u-expn}) and (\ref{e:recur-rel}). This leaves one undetermined constant, say $a_0^+$, which is an overall factor in the eigenvector. Although $a_0^\pm$ are equal, (\ref{a-3-1-ext-bound}) shows that $a_3^\pm$ (as well as $a_{2n+1}^\pm$ for $n > 1$) are generally unequal since $\G(\pm 1)$ are generally distinct. Consequently, while $\tl w, \tl w'$ and $\tl w''$ are continuous, $\tl w'''$ is generally discontinuous at the critical layer.

%-------------
\iffalse
Therefore, we can write the BCs at $\tl y = \pm 1$ and $\tl y = \tl y_{\tl \g}$ as follows.  
	\beqs 
i \tl k U'
	\begin{pmatrix}
\underbrace{E'(\G(\tl y_{\tl \g}))}_{0} & \underbrace{O'(\G(\tl y_{\tl \g}))}_{0} \\
E'(\G(\pm L)) & O'(\G(\pm L)) 
	\end{pmatrix} 
	\begin{pmatrix}
a_0^\pm\\
a_3^\pm 
	\end{pmatrix}  = 0.
	\eeqs
The condition that the determinant of this matrix vanish is automatically satisfied, so it does not impose any condition on the eigenvalue $\tl \g$. 
\fi
%-------------

\paragraph{Remark.} The nature of the continuous spectrum depends crucially on the regularity class of the eigenfunctions. In the above patched solution, $\tl w$ is $C^2$ and $\tl v$ has a discontinuous first derivative at the critical layer. One could look for other types of patched solutions, say, by starting with the $\tl v$ equation (\ref{e:v-diff-eq-const-vortex}) and defining the critical layer $\tl y_{\tl \g}$ by the condition that the coefficient of $\tl v''$ vanishes. This will potentially lead to a different continuous spectrum given by $ \{i \tl k  \tl y_{\tl \g} \pm  i \tl k/M  | -1 < \tl y_{\tl \g} < 1 \}$ with solutions lying in other regularity classes. Unlike the previous example, notice that this spectrum depends on $M$.

Admittedly, this is an incomplete study of the continuous spectrum. It remains to (i) examine the physical relevance of the continuous spectrum we have found, (ii) study the extent to which the continuous spectrum solutions to the $\tl u, \tl v$ and $\tl w$ ODEs are related when they fail to be smooth across critical layers and (iii) compare it with solutions obtained using methods of Green's functions.

%-----------------------------
\vspace{0.5cm}
%-----------------------------

{\fl \bf Acknowledgements:} We would like to thank A Thyagaraja for useful discussions, references and critical comments on the manuscript. We are also grateful to B K Shivamoggi for his valuable comments and suggesting relevant references. We also acknowledge comments from H Senapati upon going through the manuscript. This work was supported in part by the Infosys Foundation and grants (MTR/2018/000734, CRG/2018/002040) from the Science and Engineering Research Board, Govt. of India.

\appendix

%-------------------
\section{Power series solution for \texorpdfstring{$\tl w$}{w} and its regularity}
\label{s:power-series-w}
%-----------

Here, we solve the $\tl w$ equation (\ref{w-diff-eqn}) via a power series to get expansions for $\tl w, \tl \rho, \tl u$ and $\tl v$ in powers of $\G(\tl y)$. We find that these expansions define solutions that are entire functions of $\tl y$. In particular, the eigenmodes corresponding to the discrete spectrum must be smooth functions of $\tl y$ unlike the patched solutions corresponding to the continuous spectrum discussed in \S \ref{s:cont-spect}.

The $\tl w$ equation (\ref{w-diff-eqn}) is a $2^{\rm nd}$ order linear homogenous ODE with a regular singularity at $\tl y_0 = -{\tl \g}/{i \tl k}$. Although the indices $r = 0, 3$ at $\tl y_0$ are integers, the solutions $\tl w(\tl y)$ are regular at $\tl y_0$ (i.e., there are no logarithmic terms [see Chapt. X of \cite{whittaker-watson}])
	\beq
 	\tl w(\tl y) = \sum_{n=0}^{n=\infty} a_n \bigg(\tl y+\frac{\tl \g}{i\tl k }\bigg)^n = \sum_{n=0}^{n=\infty} a_n \bigg(\frac{\G}{i\tl k }\bigg)^n
   \label{exact-soln}
	\eeq
where
	\beq
	a_1 = 0, \quad a_2 = - \frac{\tl k^2}{2} a_0, \quad a_{n+4} = \frac{\tl k^2}{(n+4)(n+1)}[a_{n+2}- M^2 a_n].
	\label{e:recur-rel}
	\eeq
The coefficients $a_0$ and $a_3$ are determined by BCs while the next few are 
	\beq
	a_4 = - \frac{\tl k^2}{8} a_0 (\tl k^2 + 2M^2),\quad 	a_5 =  \tl k^2 \frac{a_3}{10} \quad 
	\text{and} \quad
	a_6 = - \tl k^4 \frac{(\tl k^2 - 2 M^2)}{144} a_0. 
	\eeq
Since (\ref{w-diff-eqn}) has no singular points other than $\tl y_0$ and $\infty$, the above power series must have infinite radii of convergence. So $\tl w$ is entire as is $\tl \rho$ (\ref{e:vort-rho-pert}). Using (\ref{u-v-w-P}), we find that $\tl v$ and $\tl u$ also have no singularities since $a_1 = 0$ and $a_0 + 2a_2/\tl k^2 = 0$ resulting in 
	\beq
	\tl v 
% = \sum_{n=0}^{n=\infty} n a_n \left({\G}/{i\tl k U'} \right)^{n-2}
	= \sum_{0}^{\infty} \frac{n+2}{i \tl k M^2 } a_{n+2} \left({\G}/{i\tl k} \right)^n \quad \text{and} \quad
	\tl u = \sum_{0}^{\infty} \frac{1}{M^2}\left( a_{n+1}+\frac{n+3}{\tl k^2} a_{n+3} \right) \left({\G}/{i\tl k } \right)^n.
	\label{v-u-expn} 
	\eeq

%-----------------------
\section{Search algorithm based on the Fredholm alternative}
\label{s:Fredholm}
%-----------------------

Here, we describe the numerical method adopted to find the growth rate $\tl \g$ by solving Eq. (\ref{e:v-diff-eq-const-vortex}) for small perturbations to the ideal background Couette flow (\ref{e:couette-ideal-flow})
	\beq
	T_{\tl \g} \tl v = (\tl k^2 + M^2 \G^2) \tl v'' - 2 M^2 \G i \tl k  \tl v' - ((\tl k^2 + M^2 \G^2)^2 + 2 \tl k^2 M^2) \tl v = 0
	\label{e:T-gamma-v-eq-0}
	\eeq
for $|\tl y| \leq 1$ subject to impenetrable boundary conditions $\tl v(\pm 1) = 0$. This may be regarded as an unconventional eigenvalue problem for the $2^{\rm nd}$ order linear ordinary differential operator $T_{\tl \g}$, with the eigenvalue $\tl \g$ appearing parametrically via $\G = \tl \g + i \tl k \tl y$. We will use the Fredholm alternative (also called the `resolvent method') to find eigenvalues $\tl \g$ and the corresponding eigenperturbations $\tl v(\tl y)$. Subject to suitable hypotheses, the Fredholm alternative \cite{Ramm, Rudin} says that either (a) the homogeneous equation $T_{\tl \g} \tl v = 0$ has a nontrivial solution $\tl v$ or (b) $T_{\tl \g} \tl v = f$ has a solution $\tl v$ for every source function $f(\tl y)$. Thus, we search for $\tl \g$ such that $T_{\tl \g} \tl v = f$ has no solution for some $f$. In practice, for a suitable fixed $f(\tl y)$ (e.g., $f = \tl y^2 - 1$ or $\cos(\pi \tl y/2)$, although $f$ need not satisfy the BCs), we move around in the complex ${\tl \g}$ plane till we reach a point where the solution $\tl v(\tl y)$ to (\ref{e:T-gamma-v-eq-0}) becomes ill-defined. In our numerical implementation (see also \cite{douglas-thyagaraja-kim-resolvent, chandra-thyagaraja} for earlier examples), we start with a $4 \times 4$ rectangular grid of trial complex values for $\tl \g$ and solve $T_{\tl \g} \tl v = f$ (at all $\tl \g$ on the grid) for $\tl v$ and compute their norms. Then we move (and eventually shrink) the $\tl \g$ grid in the complex plane to locate a point where the norm of $\tl v$ tends to become unbounded. This allows us to find an eigenvalue $\tl \g$ with eigenfunction proportional to the corresponding $\tl v$. We find that different choices for the forcing function $f$ produce essentially the same eigenvalue $\tl \g$. What is more, there is an infinite tower of eigenvalues $\tl \g$, so we need to choose the initial location of the grid to lie in the `basin of attraction' of the desired eigenvalue. In practice, we use our analytical estimates for eigenvalues $\tl \g$ [in the small $\tl k$, small $M$ or high frequency limits (see \S \ref{s:small-k-m-CC})] to initialize the grid center in the search for eigenvalues.

%-----------------------
\section{Reduction to confluent hypergeometric equation}
\label{s:hyper-geo}
%-----------------------

Upon making the transformation  
	\beq
	W = (\tl k/\G)^{1/2} \tl w  \quad \text{and} \quad Y =  (M/\tl k) \G^2
	\eeq
where $\G = \tl \g+i \tl k \tl y$, equation (\ref{w-diff-eqn}) reduces to Whittaker's equation \cite{Abram-Stegun}
	\beq
	\frac{d^2 W}{d Y^2} + \left(-\frac{1}{4} + \frac{\nu}{Y}+ \frac{\frac{1}{4}-\mu^2}{Y^2} \right)W = 0 \quad \text{where} \quad
\nu = \frac{\tl k}{4 M} \quad \text{and} \quad \mu = \frac{3}{4}.
	\eeq
Solutions can be written as $W(Y) = c_1 M_{\nu,\mu}(Y)+ c_2 M_{\nu,-\mu}(Y)$ where 
	\beq
	M_{\nu,\mu}(Y) = e^{-Y/2}Y^{\mu+1/2}\: _1F_1(1/2 + \mu-\nu, 1+2\mu, Y).
	\eeq
Here $_1 F_1$ is Kummer's confluent hypergeometric function. Impenetrable boundary conditions lead to
	\beq
	\tl k W(Y_{\pm 1}) + 4 
M (\tl \g \pm i \tl k)^2 W'(Y_{\pm 1})  = 0 
	\label{e:bc-hyp}
	\eeq
where $Y_{\pm 1} = (M/\tl k) (\tl \g \pm i \tl k)^2$. In principle, one could solve this to determine the allowed growth rates $\tl \g (M, \tl k)$. However, we do not pursue such a direct analytical approach in this paper, opting instead to obtain a quantitative picture of linear instabilities by combining stability theorems, growth rate bounds, limiting cases, series solutions and numerical spectra. We hope to return elsewhere to an examination of the spectrum of this eigenvalue problem (and its parametric dependence on $\tl k$ and $M$) using complex analytic methods (see \cite{coddington-levinson} and also \cite{Glat-W}). In \S \ref{s:small-k-m-CC} we will solve (\ref{w-diff-eqn}) in limiting cases analytically and treat the general case numerically in \S \ref{s:lc-regime}.

%-----------------------

\end{document}